\documentclass[lettersize,journal]{IEEEtran}
\usepackage{amsmath,amsfonts}
\usepackage[psamsfonts]{amssymb}
\usepackage{amsxtra}
\usepackage[linesnumbered,boxed,ruled,commentsnumbered]{algorithm2e}
\usepackage{array}
\usepackage[caption=false,font=footnotesize,labelfont=sf,textfont=sf]{subfig}
\usepackage{color}
\usepackage{threeparttable}
\usepackage{textcomp}
\usepackage{stfloats}
\usepackage{url}
\usepackage{verbatim}
\usepackage{graphicx}
\usepackage{cite}
\usepackage{booktabs}
\usepackage{makecell}
\usepackage{epstopdf}
\usepackage{multirow}
\usepackage{graphicx}

\usepackage{comment}

\graphicspath{{./figure/} }
\DeclareGraphicsExtensions{.pdf,.jpeg,.png}

\usepackage[colorinlistoftodos]{todonotes}

\def\defeq{\stackrel{\triangle}{=}}
\newcommand{\fontpersosmall}[1]{\textsf{\scriptsize{#1}}}
\newcommand{\fontperso}[1]{\textsf{\small{#1}}}
\makeatletter
\newcommand*{\bigcdot}{}
\DeclareRobustCommand*{\bigcdot}{%
  \mathbin{\mathpalette\bigcdot@{}}%
}
\newcommand*{\bigcdot@scalefactor}{0.8}
\newcommand*{\bigcdot@widthfactor}{1.5}
\newcommand*{\bigcdot@}[2]{%
  \sbox0{$#1\vcenter{}$}
  \sbox2{$#1\cdot\m@th$}%
  \hbox to \bigcdot@widthfactor\wd2{%
    \hfil
    \raise\ht0\hbox{%
      \scalebox{\bigcdot@scalefactor}{%
        \lower\ht0\hbox{$#1\bullet\m@th$}%
      }%
    }%
    \hfil
  }%
}
\makeatother
\begin{document}

\title{Determined Blind Source Separation with Sinkhorn Divergence-based Optimal Allocation of the Source Power}

\author{Jianyu Wang, Shanzheng Guan, Nicolas Dobigeon,~\IEEEmembership{Senior Member,~IEEE}, Jingdong Chen,~\IEEEmembership{Fellow,~IEEE}
\thanks{J. Wang, S. Guan, J. Chen are with the Center of Intelligent Acoustics and
Immersive Communications, Northwestern Polytechnical University, China (email: \{alexwang96, gshanzheng\}@mail.nwpu.edu.cn, jingdongchen@ieee.org).}
\thanks{N. Dobigeon is with University of Toulouse, IRIT/INP-ENSEEIHT, 2 rue Charles Camichel, BP 7122, 31071 Toulouse Cedex 7, France (e-mail: nicolas.dobigeon@enseeiht.fr).}
\thanks{This work was supported in part by the Artificial Natural Intelligence Toulouse Institute (ANITI, ANR-19-PI3A-0004).}}

\maketitle

\begin{abstract}
Blind source separation (BSS) refers to the process of recovering multiple source signals from observations recorded by an array of sensors. Common approaches to BSS, including independent vector analysis (IVA), and independent low-rank matrix analysis (ILRMA), {{typically rely on second-order models to capture the statistical independence of source signals for separation. However, these methods generally do not account for the implicit structural information across frequency bands, which may lead to model mismatches between the assumed source distributions and the distributions of the separated source signals estimated from the observed mixtures.}}
To tackle these limitations, this paper shows that conventional approaches such as IVA and ILRMA can easily be leveraged by the Sinkhorn divergence, {{incorporating an optimal transport (OT) framework to adaptively correct source variance estimates.}}
This allows for the recovery of the source distribution while modeling the inter-band signal dependence and reallocating source power across bands. As a result, enhanced versions of these algorithms are developed, integrating a Sinkhorn iterative scheme into their standard implementations. Extensive simulations demonstrate that the proposed methods consistently enhance BSS performance.
\end{abstract}


\section{Introduction}\label{sec:intro}
\label{sect-intro} \IEEEPARstart{T}{he} objective of multichannel blind source separation (MBSS) is to extract distinct source components from their mixtures observed by an array of sensors \cite{1,2,3}. MBSS methods can be generally categorized into two groups, depending on the relationship between the number of sensors and sources. Determined/overdetermined methods assume that the number of sensors equals or exceeds the number of sources \cite{4}. Conversely  underdetermined methods account for the case of  fewer sensors than sources \cite{5}. Although both families have received significant attention in the literature, the underdetermined scenario poses more technical difficulties, often resulting in separated signals with increased distortion compared to the determined cases. This work focuses on the determined case to achieve sound source separation in reverberant environments with minimized distortion.

Separating sound sources in reverberant environments is typically formulated as a convolutive MBSS problem in the frequency domain. A large number of methods and algorithms have been developed to perform this task, including frequency-domain independent component analysis (ICA) \cite{6,7}, independent vector analysis (IVA) \cite{8,9,10}, and independent low-rank matrix analysis (ILRMA) \cite{4,10,11}. All these approaches rely on a rank-1 spatial model and explore the independence of source signals to achieve source signal recovery. However ICA generally suffers from the well-known issue of permutation \cite{7}. In contrast, IVA addresses the permutation issue by considering statistical dependencies among demixed signals across different frequency bands \cite{9}. ILRMA further explores the low-rank spectral structure of sources in addition to the statistical independence among sources by recasting the problem into the framework of nonnegative matrix factorization (NMF). Several improved versions of ILRMA have been also proposed, including Student's $t$-distribution-based ILRMA \cite{14}, Student's $t$-distribution-based MNMF \cite{13}, generalized-Gaussian-distribution-based ILRMA (GGD-ILRMA) \cite{15,33} and heavy-tailed FastMNMF \cite{22}. Despite these advances, some opportunities should be taken to further improve the source and spatial models. In particular, significant efforts have been made to refine the source modeling, such as the frequency-wise sparse regularization \cite{18}, minimum-volume prior \cite{16,17}, time-frequency mask \cite{19}, near-field source modeling \cite{21}, high-resolution NMF \cite{31}, source spectral modeling \cite{32}, probabilistic modeling \cite{36}, Gaussian process regression \cite{37}, iterative projection \cite{38}, and inter-frame modeling \cite{sawada2023multi}. Conversely, other contributions have been devoted to improving the spatial model in MBSS, exploring direction-of-arrival (DOA) kernels \cite{26},\cite{23}, confining the spatial covariance matrices to be
jointly-diagonalizable \cite{20}, combining MBSS with multichannel {{Wiener}} filter \cite{28}, energy constraining \cite{34}, considering BSS and reverberation simultaneously \cite{nakatani2022switching}, \cite{sekiguchi2022autoregressive}, to name but a few.

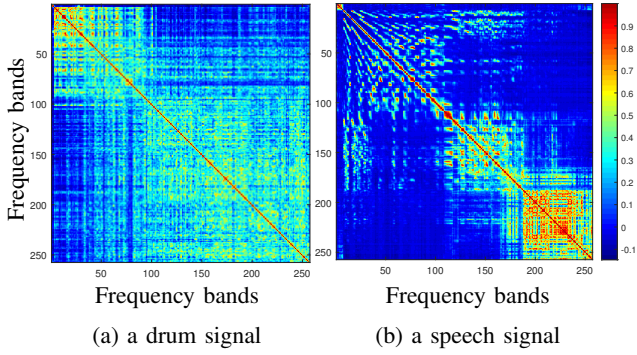
\begin{figure}[t]
\begin{tikzpicture}
\node[inner sep=0pt] (oSNR) at (-45pt,0pt)
    {\includegraphics[width=43mm]{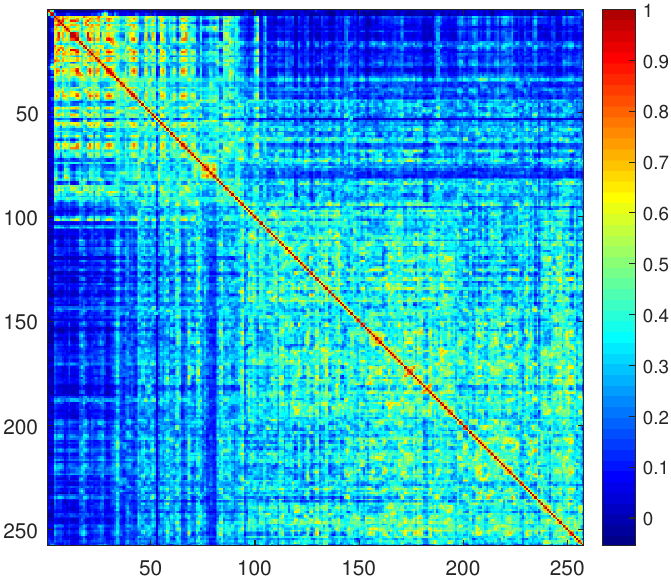}};
\draw  node[] at (-50pt,-60pt) {\small Frequency bands};
\draw  node[] at (-50pt,-75pt) {\small (a) a drum signal};
\draw node[rotate=90,  anchor=center] at (-110pt, 0pt) {\small Frequency bands};

\node[inner sep=0pt] (vsd) at (63pt, 0pt)
    {\includegraphics[width=43mm]{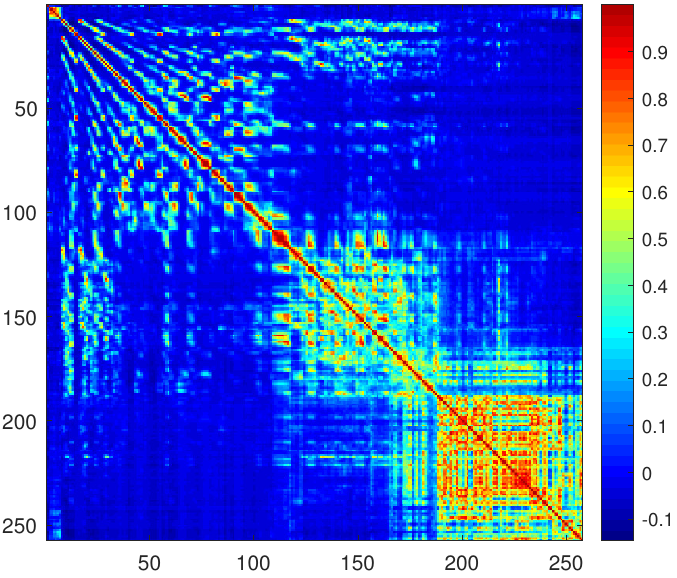}};
\draw  node[] at (60pt,-60pt) {\small Frequency bands};
\draw  node[] at (60pt,-75pt) {\small (b) a speech signal};
\end{tikzpicture}
\vspace{-0.3cm}
\caption{Illustration of inter-band correlation of: (a) a drum signal, and (b) a speech signal.}
\label{fig-interbandcorrelation}
\end{figure}

{{Although many existing multichannel BSS (MBSS) methods, such as IVA and ILRMA, leverage dependencies across different short-time Fourier transform (STFT) bands, they typically do so through implicit models rather than by explicitly modeling inter-band correlations. For instance, IVA assumes a joint distribution across all frequency bands, while ILRMA utilizes non-negative matrix factorization (NMF) to represent the spectral structure. However, these implicit approaches may not fully capture the actual spectral correlations present in real-world speech signals. In the field of speech enhancement and noise reduction, studies have demonstrated significant correlations between spectral components in adjacent STFT bands, and leveraging these inter-band correlations has been shown to improve performance \cite{5742773}. Therefore, investigating how to explicitly model inter-band dependencies to enhance MBSS performance remains an important question, which we address in this work. }}
In this study, we adopt the NMF framework for MBSS and focus on explicitly modeling the dependencies among magnitude square spectra, rather than the complex STFT spectra used in noise reduction \cite{54, chen2012single}. We refer to these dependencies as inter-band power correlations, even though magnitude square spectral components are not strictly power spectral densities. This terminology simplifies the discussion while capturing the essential relationships. Figure~\ref{fig-interbandcorrelation} illustrates the inter-band power correlation coefficients for two types of signals \cite{Liang2014}, revealing strong dependencies between magnitude square spectral components from various STFT bands. These correlations persist even between distant frequency bands, which contrasts with findings in noise reduction studies.

{{Another critical limitation of existing MBSS methods is the model mismatch arising from differences between the assumed and actual source distributions. Typically, source separation models assume that source signals follow a zero-mean complex Gaussian distribution. However, observed signals often exhibit complex, non-Gaussian distributions. After estimation with the demixing matrix, this results in inconsistencies between the distributions of the separated sources and the assumed source distributions, highlighting a model mismatch.

Our method addresses model mismatch stemming from differences between assumed and actual source distributions by leveraging the optimal transport (OT) framework \cite{peyre2019computational}, specifically through Sinkhorn divergence. In typical source separation models, source signals are assumed to follow a zero-mean complex Gaussian distribution, while observed signals often exhibit complex, non-Gaussian characteristics. This inconsistency results in distributional mismatches between the separated sources and the assumed source models, which our approach minimizes.

To approximate these two distributions and reduce model mismatch, we utilize Sinkhorn divergence—a regularized version of the Wasserstein (OT) distance \cite{39,41,42,43,44,scheibler2019independent}, also known as Earth mover’s distance. The Wasserstein distance has been widely used to gauge the similarity between distributions and capture spectral spread across frequency bands \cite{45}. However, in its standard form, it introduces significant computational complexity, particularly in high-dimensional settings. By adopting Sinkhorn divergence, we can retain the benefits of distribution alignment while achieving computational efficiency. This alignment allows the estimated variance to better capture the overall signal distribution pattern, leveraging both inter-band dependency and distributional discrepancies for more accurate source separation.}}
To overcome this limitation, this work proposes to consider the Sinkhorn divergence,  a regularized version of the Wasserstein distance which offers the possibility of resorting to computationally cheaper optimization strategies. More precisely, this work proposes to extend two popular MBSS approaches, namely IVA and ILRMA, to show that they can be easily granted with a Sinkhorn divergence-based re-allocation of the source power.

{{Unlike conventional IVA and ILRMA, the proposed algorithms utilize the Sinkhorn divergence, which effectively combines both vertical divergence and horizontal distance. The vertical divergence compares magnitude square spectrograms at individual frequency bands (point-wise comparison along the same frequency), while the horizontal distance captures how spectral energy is distributed across different frequency bands (overall comparison across frequencies). By incorporating both these aspects, Sinkhorn divergence ensures a more comprehensive alignment of the estimated and true source distributions, leading to improved BSS performance.}}

\section{Conventional formulations of IVA and ILRMA}\label{sec:conventional}

This work considers an acoustic application scenario characterized by $N$ sources and  $M$ microphones.
Let $h_{mn}(\cdot)$ denote the acoustic impulse response associated with the $n$th source and
the $m$th microphone. The length of the impulse response is $L$ samples.
By neglecting the effect of background noise,  the signal recorded by the $m$th microphone at time instant $\tau$ can
be written as
\begin{align}
\label{MixSysT}
 {x}_m(\tau) = \sum_{n=1}^N \sum_{l=0}^{L-1} h_{mn}(l) s_n(\tau - l), \quad m = 1,\ldots,M,
\end{align}
where $s_n(\tau)$ denotes the $n$th source
signal. This convolution measurement model can be expressed {{approximately}} into the short-time-Fourier-transform
(STFT) domain as
\begin{align}\label{MixSysSTFT}
 x_m(f,t) = \sum_n^N [\mathbf{H}(f)]_{mn} s_n(f,t), \quad m = 1,\ldots,M,
\end{align}
where $x_{m}(f,t)$ and $s_{n}(f,t)$ denote STFTs of ${x}_m(\tau)$ and $s_n(\tau)$, respectively, at the frequency-bin index $f\in \left\{1,\ldots,F\right\}$ and time-frame index $t\in \left\{1,\ldots,T\right\}$, with $F$ and $T$ denoting the total numbers of frequency bins and time frames. In \eqref{MixSysSTFT},
the matrix $\mathbf{H}(f)$ is of size $M \times N$, whose $m$th-column and
$n$th-row element corresponds to the STFT of $h_{mn}$ at the $f$th frequency
bin.

For simplicity, we assume in this work that $M=N$ (however, the algorithm developed here can be readily extended to the case where  $M>N$ {{\cite{scheibler2019independent}}}) and $\mathbf{H}(f)$ matrix has full rank. Thus, in the following formulations and derivations, we shall replace $M$ by $N$. The objective of MBSS is then to estimate the sources
$\left\{s_n(f,t)\right\}_{n,f,t}$ given the observed signals
$\left\{ x_n(f,t) \right\}_{n,f,t}$. By introducing
the measurement vector $\mathbf{x}(f,t)$ as
\begin{align}
\label{observector}
 \mathbf{x}(f,t) \defeq \left[\begin{array}{cccc} {x}_1(f,t) & {x}_2(f,t) & \cdots & {x}_N(f,t) \end{array}\right]^\intercal,
\end{align}
the problem can be formulated as recovering a reconstructed source $\mathbf{y}(f,t)$ of the unknown source vector $\mathbf{s}(f,t)$ such that
\begin{align}\label{SourceEst}
 \mathbf{y}(f,t) = \mathbf{D}(f) \mathbf{x}(f,t),
\end{align}
where both $\mathbf{y}(f,t)$ and $\mathbf{s}(f,t)$ are defined analogously to $\mathbf{x}(f,t)$,
\begin{align}
\mathbf{D}(f) &\defeq
\left[\begin{array}{cccc} \mathbf{d}_{1}(f) & \mathbf{d}_{2}(f) & \cdots & \mathbf{d}_{N}(f) \end{array}\right]
\end{align}
is referred to as the demixing matrix and has dimensions $N \times N$ with
\begin{align}
\label{observector2}
\mathbf{d}_n(f) &\defeq
\left[\begin{array}{cccc} {d}_{n,1}(f) & {d}_{n,2}(f) & \cdots & {d}_{n,N}(f) \end{array}\right]^\intercal.
\end{align}

\subsection{Independent vector analysis}

With a slight abuse of notations, let us denote the whole set of measurements, reconstructed sources and demixing matrices as, respectively,
\begin{align}
\mathcal{X} &\defeq
\begin{bmatrix}
\mathbf{x}(1,1) & \dots & \mathbf{x}(1,T)\\
\vdots &\ddots & \vdots \\
\mathbf{x}(F,1)& \dots & \mathbf{x}(F,T)
\end{bmatrix}\in\mathbb{C}^{N\times F\times T}, \quad \\
\mathcal{Y} &\defeq
\begin{bmatrix}
\mathbf{y}(1,1) & \dots & \mathbf{y}(1,T)\\
\vdots &\ddots & \vdots \\
\mathbf{y}(F,1)& \dots & \mathbf{y}(F,T)
\end{bmatrix}\in\mathbb{C}^{N\times F\times T}, \quad
\end{align}
and
\begin{align}
\mathcal{D} &\defeq
\begin{bmatrix}
\mathbf{d}_1(1) & \dots & \mathbf{d}_1(F)\\
\vdots &\ddots & \vdots \\
\mathbf{d}_N(1)& \dots & \mathbf{d}_N(F)
\end{bmatrix}\in\mathbb{C}^{N\times {{N\times F}}},
\end{align}

Within the Bayesian framework, the joint posterior distribution of the unkown demixing
matrices and reconstructed sources is given by $p(\mathcal{D}, \mathcal{Y}|\mathcal{X})  \propto  p(\mathcal{X}|\mathcal{D}  , \mathcal{Y}) p(\mathcal{Y}|\mathcal{D}) p(\mathcal{D})$. IVA elaborates on this posterior distribution by forming a maximum a posteriori (MAP) estimator of the demixing matrix $\mathcal{D}$, which is derived by solving the following optimization problem:
\begin{equation}\label{PosterioriIVAproblem}
   \mathop{\min}\limits_{\mathcal{D}}~ \left\{- \log   p(\mathcal{D}|\mathcal{X})\right\},
\end{equation}
where $p(\mathcal{D}|\mathcal{X}) \propto p(\mathcal{X}|\mathcal{D})  p(\mathcal{D})$ and
\begin{equation}\label{PosterioriIVAproblem2}
   p(\mathcal{X}|\mathcal{D}) = \int p(\mathcal{X}|\mathcal{D},\mathcal{Y}) p(\mathcal{Y}|\mathcal{D}) d\mathcal{Y}.
\end{equation}
Besides, IVA introduces key assumptions regarding the prior independence of the sources
and prior independence across time frames. Specifically, it is assumed that the conditional
probability distribution of $\mathcal{Y}$ can be factorized as
\begin{align}\label{PosteriorEst1}
\begin{split}
 p(\mathcal{Y}|\mathcal{D}) = \prod_{n=1}^{N} \prod_{t=1}^{T} p\left(\mathbf{y}_n(:,t)|\mathcal{D}\right),
\end{split}
\end{align}
where $\mathbf{y}_n(:,t) \in \mathbb{C}^F$.
Moreover, it is assumed that the errors between the observed signals and their
reconstructed counterparts are independent and identically distributed so the
full likelihood function can be written as
\begin{align}
\label{PosteriorEst3}
 p(\mathcal{X}|\mathcal{D},\mathcal{Y}) = \prod_{n=1}^{N} \prod_{f=1}^{F} \delta \left[ \mathbf{x}_n(f,:) - \mathbf{D}^{-1}(f)\mathbf{y}_n(f,:) \right]
\end{align}
{{where $\delta[\cdot]$ denotes the Delta function,}}
and the marginal likelihood can be written as
\begin{align}\label{PosteriorEst4}
\begin{split}
 p(\mathcal{D}|\mathcal{X}) \propto p(\mathcal{D}) \prod_{f=1}^{F} \left| \det\mathbf{D}(f) \right|^{2N} \prod_{n=1}^{N} \prod_{t=1}^{T} p\left(\mathbf{y}_n(:,t)|\mathcal{D}\right).
\end{split}
\end{align}
{{Since directly estimating the distribution of $\mathbf{y}_n(:,t)$ is highly challenging due to the complexity and non-Gaussian nature of the observed mixed signals, many literatures \cite{9,10,11} instead replace $p\left(\mathbf{y}_n(:,t)|\mathcal{D}\right)$ with the hypothesis distribution $p\left(\mathbf{s}_n(:,t)\right)$ of the source signal
\begin{align}\label{PosteriorEst5}
\begin{split}
 p(\mathcal{D}|\mathcal{X}) \propto p(\mathcal{D}) \prod_{f=1}^{F} \left| \det\mathbf{D}(f) \right|^{2N} \prod_{n=1}^{N} \prod_{t=1}^{T} p\left(\mathbf{s}_n(:,t)\right).
\end{split}
\end{align}
}}
Finally, the optimization problem in \eqref{PosterioriIVAproblem} is transformed into the following (marginal) MAP estimator:
\begin{align}\label{PosterioriIVA}
\hat{\mathcal{D}}_{\mathrm{MAP}} = & \arg \mathop{\min}\limits_{\mathcal{D}}~\left\{- \log p(\mathcal{D}) - 2N \sum_f \log \left| \det \mathbf{D}(f) \right| \right. \nonumber \\
 & \phantom{\arg \mathop{\max}\limits_{\mathcal{D}}}~~~~ \left. + \sum_{n,t} G\left[\mathbf{s}_n(:,t)\right] \right\},
\end{align}
where $G\left[\mathbf{s}_n(:,t)\right] = - \log
p\left(\mathbf{s}_n(:,t)\right)$ denotes a contrast function, {{and $p(\mathcal{D})$ denotes the prior distribution of the demixing matrix (That's not the focus of our research)}}. A popular
choice for this prior source model is the spherical super Gaussian
distribution {{\cite{9, 10}}}, i.e.,
\begin{align}\label{SSGSource1}
 p\left(\mathbf{s}_n(:,t)\right)  \propto  \exp  \left( -  \| \mathbf{s}_n(:,t) \|_2  \right).
\end{align}
Another typical choice is a time-varying variance Gaussian distribution \cite{4}:
\begin{align}\label{SSGSource2}
 p\left(\mathbf{s}_n(:,t)\right) \propto \exp  \left( - \frac{ \| \mathbf{s}_n(:,t) \|_2^2 \!}{\sigma_{n,t}^2}  \right),
\end{align}
where $\sigma_{n,t}$ denotes the variance of the Gaussian distribution. Interestingly, {{the equation \eqref{SSGSource2} assumes a time-varying variance Gaussian distribution with frequency-independent variance $\sigma_{n,t}$.
However, the model should account for frequency dependence as well. Specifically, the variance should vary both with time and frequency, which is captured by introducing a diagonal covariance matrix in equation \eqref{SSGSource3}.}} The source model \eqref{SSGSource2} explicitly specifies a contrast function known as the Itakura-Saito (IS) divergence
 \cite{fevotte2009nonnegative}, i.e.,
\begin{align}\label{SSGSource4}
 G \left[  \mathbf{s}_n(:,t) ; \boldsymbol{\sigma}_{n,t} \right]  =  \sum_{f}  \left(  \frac{\left| s_n(f,t) \right|^2}{ \sigma_{n,f,t}^2 } +  \log \sigma_{n,f,t}^2  \right).
\end{align}
{{where $\boldsymbol{\sigma}^{\bigcdot 2}_{n,t} \defeq
\left[ \begin{array}{cccc}
{\sigma}^2_{n,1,t} & {\sigma}^2_{n,2,t} & \cdots & {\sigma}^2_{n,F,t}
\end{array} \right]^\intercal.$\footnote{With a slight abuse of notation, $ ^{\bigcdot}$ indicates that the following operation is performed element-wise.}}}
The optimization problem presented in (\ref{PosterioriIVA}) can be tackled using different approaches. One particularly efficient method, proposed in \cite{10}, employs a majorize-minimization technique, eliminating the necessity to adjust a step size parameter.

\subsection{ILRMA}
{{In contrast to IVA, ILRMA recasts the estimation problem into an NMF framework \cite{lee2000algorithms}, which implicitly captures dependencies across frequency bands through the low-rank structure of the source model. It employs a time-varying variance complex Gaussian distribution to contruct}}
the source model, i.e.,
\begin{align}\label{SSGSource3}
p\left[\mathbf{s}_n(:,t)\right]= & \prod_{f=1}^{F} \mathcal{N}_{\mathbb{C}} \left[ s_n(f,t) | 0, \sigma_{n,f,t}^2 \right] \nonumber \\
= & \prod_{f=1}^{F} \frac{1}{\pi\sigma_{n,f,t}^2} \exp \left[ - \frac{\left| s_n(f,t) \right|^2}{\sigma_{n,f,t}^2} \right],
\end{align}
where $\sigma_{n,f,t}^2 \defeq \mathbb{E}\left[\left| s_n(f,t) \right|^2\right]$. The matrix for the time-varying variances for all the time frames and
frequency bins is
{{
\begin{align}\label{SourceModelVariance}
 {\sigma}_{n,f,t}^2 &= \sum_{k=1}^K u_{n,f,k} v_{n,k,t}.
\end{align}
where $u_{n,f,k}$ and $v_{n,k,t}$}}
denote the basis and activation matrices of the source model, respectively, and $K$ is the factorization order. It is noteworthy that similar to the model in \eqref{SSGSource4}, the source model described in \eqref{SourceModelVariance} also explicitly defines a contrast function known as the IS divergence.

{{
\subsection{Analysis}
In the standard formulation, as shown in equation \eqref{PosteriorEst4}, it is typically assumed that $p\left(\mathbf{y}_n(:,t)|\mathcal{D}\right) = p\left(\mathbf{s}_n(:,t)\right)$, allowing the observed signals to be modeled by the assumed source distribution $p\left(\mathbf{s}_n(:,t)\right)$. This simplification reduces computational complexity by avoiding the need to directly estimate the complex, non-Gaussian distribution of the observed signals. However, in practical scenarios, the true distribution of observed signals often diverges from this assumption due to factors like noise and reverberation, resulting in model mismatches that reduce separation accuracy.

To address these mismatches, our method employs Sinkhorn divergence, a regularized version of the Wasserstein (OT) distance, which corrects the discrepancies between the assumed and estimated source distributions through an optimal transport matrix $\mathbf{Q}$ (defined in the next section). Unlike traditional Wasserstein distance, Sinkhorn divergence provides an efficient alignment between distributions while minimizing computational complexity. By aligning the estimated source power with the variance factor of the assumed source distribution, Sinkhorn divergence allows the model to more accurately capture the signal’s overall distribution pattern. This approach reduces model mismatch error, enabling more reliable and accurate source separation in real-world conditions.
}}

\section{Sinkhorn divergence based BSS}
\label{section3}
\subsection{Optimal mapping of the source parameters}\label{subec:source}

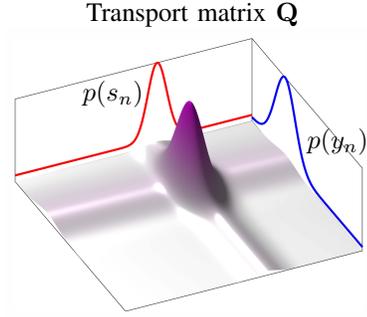
\begin{figure}
\centering
\begin{tikzpicture}
\node[inner sep=0pt] (oSNR) at (30pt,0pt)
    {\includegraphics[width=50mm]{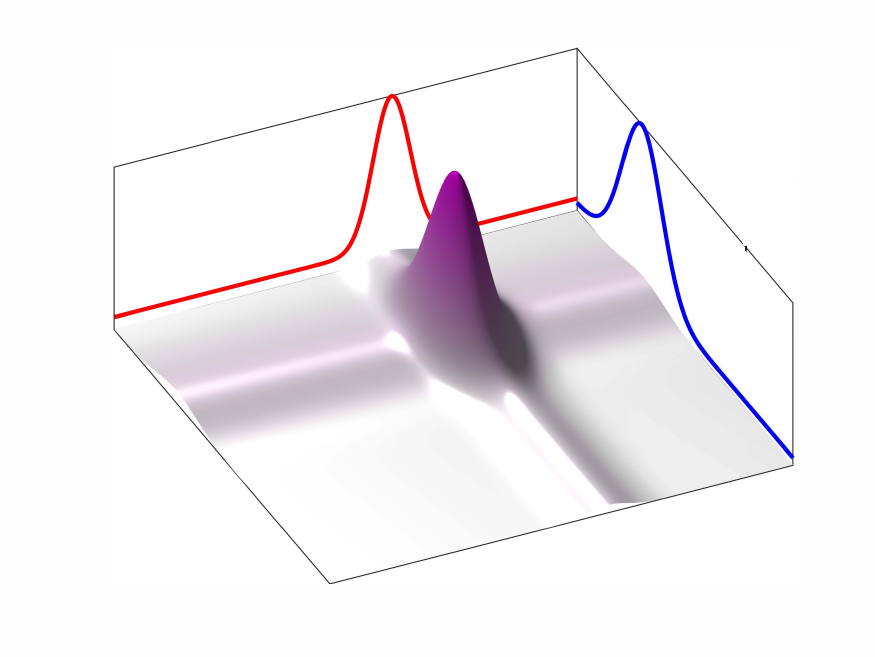}};
\draw  node[] at (0pt,30pt) {$p(s_n)$};
\draw  node[] at (85pt,12pt) {$p(y_n)$};
\draw  node[] at (30pt,60pt) {Transport matrix $\mathbf{Q}$};
\end{tikzpicture}
\vspace{-0.3cm}
\caption{Visualization of the transport matrix $\mathbf{Q}$ {{linking the empirical magnitude square spectrum of $p(y_n)$ to the variance parameters of the source $p(s_n)$.}}
\label{fig3}}
\end{figure}

As described in Section \ref{sec:intro}, the objective of this study is to utilize the OT framework to align the estimated source distribution with the true source model distribution, taking into consideration inter-frequency dependencies. This objective is framed as minimizing the transport path from the variance parameters of the estimated signal distribution to the parameters characterizing the hypothetical source distribution. To alleviate the computational load associated with conventional OT formulations, we employ the 
{{Sinkhorn distance}}\footnote{{The Wasserstein distance is the fundamental metric of optimal transport, calculating the minimal "transport cost" between two probability distributions. The Sinkhorn distance introduces entropy regularization to the Wasserstein distance and optimizes the transport plan using Sinkhorn iterations. The Sinkhorn divergence further regularizes the Wasserstein distance with entropy and incorporates self-similarity alignment terms. The dual-Sinkhorn divergence is based on the dual formulation. For a detailed explanation, please refer to \cite{peyre2019computational}.}}, which is defined \cite{44}:
\begin{align}\label{SourcePrior4bis}
\begin{split}
 {{\mathcal{S}_{\frac{1}{\lambda}}^1}} \! \left( \mathbf{z}_1, \mathbf{z}_2 \right) = \mathop{\min}\limits_{\mathbf{Q} \in \Pi\left( \mathbf{z}_1, \mathbf{z}_2 \right)} \left[\left\langle \mathbf{Q},\mathbf{C} \right\rangle - \frac{1}{\lambda} H(\mathbf{Q}) \right],
\end{split}
\end{align}
where {{the angle brackets notation $\left\langle \cdot, \cdot \right\rangle$ represents the Frobenius inner product, and}}
\begin{align}
H(\mathbf{Q}) = -\sum_{f,f^\prime} \left[\mathbf{Q}\right]_{f,f^\prime} \log \left[\mathbf{Q}\right]_{f,f^\prime},
\end{align}
$\mathbf{C}\in\mathbb{R}^{F\times F}$ denotes the quadratic cost matrix associated with the Euclidean distance {{between the frequency components of magnitude square spectrum in the empirical source distribution and the hypothetical source distribution}}, and the set of transport plans $\Pi\left( \mathbf{z}_1,\mathbf{z}_2 \right)$ are defined as
\begin{align}\label{SourcePrior4_1}
\begin{split}
 \Pi\left( \mathbf{z}_1,\mathbf{z}_2 \right) = \left\{ \mathbf{Q}\in\mathbb{R}^{F\times F}_+: \mathbf{Q}\mathbf{1} = \mathbf{z}_1, \mathbf{Q}^\intercal\mathbf{1} = \mathbf{z}_2 \right\}
\end{split}
\end{align}
with $\mathbf{1}$ denoting the
all-one vector. When the variables $\mathbf{z}_1$ and $\mathbf{z}_2$ are defined as the {{magnitude square spectrum between the empirical source and the hypothetical true source}},
minimizing \eqref{SourcePrior4bis} provides the optimal transport plan matrix {{$\mathbf{Q}$}} mapping the two distributions {{between the estimated source and the hypothetical true source}}, as schematically illustrated in Fig.~\ref{fig3}.

To model the dependency between frequency bands under the assumption that the sources follow a zero-mean complex Gaussian distribution, {{leveraging Sinkhorn divergence to model the variance relationships between estimated source power and the variance parameters of the source model}}, is followed for the sets of band-by-band source energy $\boldsymbol{\sigma}^{\bigcdot 2}_{n,t}$ and $|\mathbf{y}_n(:,t)|^{\bigcdot 2}$, respectively. This leads to the search for the set of optimal transport matrices $\hat{\mathbf{Q}}_{n,t}$ associated with the $n$th source at $t$th time frame. When the reconstructed sources
$\mathbf{y}_n(:,t)$ are inferred from {{the demixing matrix estimated through IVA or ILRMA, which provides the separated source signals,}} the energy levels of the hypothetical true sources
$\mathbf{s}(:,t)$ can be updated after estimating the optimal transport matrices. Specifically, let assume that the reconstructed sources $\mathbf{y}_n(:,t)$ are known, the optimal mapping with the $n$th hypothetical source parameters at time $t$ is given by
\begin{equation} \label{eq:optimal_mapping_sigma]}
\hat{\boldsymbol{\sigma}}^{\bigcdot 2}_{n,t} = \hat{\mathbf{Q}}_{n,t} \mathbf{1},
\end{equation}
where
\begin{equation}\label{eq:optimal_mapping}
  \hat{\mathbf{Q}}_{n,t} = \arg  \min_{\mathbf{Q}_{n,t} \geq 0} \mathcal{S}_{\frac{1}{\lambda},\gamma}^2\left( \boldsymbol{\sigma}^{\bigcdot 2}_{n,t} \Big| |\mathbf{y}_n(:,t)_{\bigcdot}|^{2} \right).
\end{equation}
{{Here, $\hat{\mathbf{Q}}_{n,t} \geq 0$ indicates that $\hat{\mathbf{Q}}_{n,t}$ is a non-negative matrix, and all elements of the matrix $\hat{\mathbf{Q}}_{n,t}$ are non-negative.}}
In \eqref{eq:optimal_mapping}, the objective function $\mathcal{S}_{\gamma,\frac{1}{\lambda}}^2(\cdot|\cdot)$ is defined from a variational counterpart of the Sinkhorn divergence \cite{peyre2019computational} given in \eqref{SourcePrior4bis}, i.e.,
\begin{align}\label{eq:optimal_mapping2}
 & \mathcal{S}_{\frac{1}{\lambda},\gamma}^2\left( \boldsymbol{\sigma}^{\bigcdot 2}_{n,t} \Big| |\mathbf{y}_n(:,t)_{\bigcdot}|^{2} \right)  =  \left\langle \mathbf{Q}_{n,t},\mathbf{C}_{n,t} \right\rangle   -  \frac{1}{\lambda} \! H\left(\mathbf{Q}_{n,t}\right)   \nonumber  \\
 &  \ +  \gamma\left[ \mathcal{L}_{\varphi} \left(\mathbf{Q}_{n,t}\mathbf{1}\Big|\boldsymbol{\sigma}^{\bigcdot 2}_{n,t}\right) +  {\mathcal{L}_{\varphi}} \left(\mathbf{Q}_{n,t}^\intercal\mathbf{1}\Big||\mathbf{y}_n(:,t)_{\bigcdot}|^{2}\right) \right],
\end{align}
where
\begin{align}
   \left[\mathbf{C}_{n,t}\right]_{f,f^\prime} &= \left| \left| {y}_n(f,t) \right|^{2} - {\sigma}^{ 2}_{n,f^\prime,t} \right|,  \\
   \left\langle \mathbf{Q}_{n,t},\mathbf{C}_{n,t} \right\rangle &= \sum_{f\!,f^\prime} \left[\mathbf{Q}_{n,t}\right]_{f\!,f^\prime} \! \bigcdot \! \left[\mathbf{C}_{n,t}\right]_{f\!,f^\prime},
\end{align}
and where $\mathcal{L}_\varphi\left(\cdot|\cdot\right)$ is a distance measure, which is chosen as the KL divergence in this work to enforce the marginal alignment constraint.
{{The objective function in \eqref{eq:optimal_mapping2} consists of four terms. The first two terms form the Wasserstein distance, where the transport matrix $\mathbf{Q}_{n,t}$ constrains the joint distribution between the estimated source power spectrum and the hypothesized source variance, ensuring alignment in terms of power distribution. The third and fourth terms of \eqref{eq:optimal_mapping2} enforce marginal alignment constraints, using the KL divergence to match the marginal distributions of the estimated and modeled sources. Together, these terms define the Sinkhorn divergence, aligning the estimated source distribution with the modeled one while minimizing regularization bias.}}

\begin{algorithm}[t]
\label{algo:OT}
\small
\setlength{\tabcolsep}{0.1mm}{
  \SetKwInOut{Input}{\textbf{Input}}   
  \caption{Computing the optimal mapping matrix $\hat{\mathbf{Q}}_{n,t}=\mathsf{OM}\left[\boldsymbol{\sigma}^{\bigcdot 2}_{n,t},\mathbf{y}_n(:,t)\right]$ between the hypothetical and reconstructed source parameters. }
  \Input{
         Reconstructed sources $\mathbf{y}_n(:,t) \in \mathbb{C}^{{{F}}}$, estimated source parameters $\boldsymbol{\sigma}^{\bigcdot 2}_{n,t}$,
         number of iterations $J_\mathrm{s}$,
         hyperparameters $\lambda$, $\gamma$ and $r$.\\
        }
    \BlankLine
        {
        {{Initialize the $\boldsymbol{\xi}_{n,t}$.}} \\ $\left[\mathbf{C}_{n,t}\right]_{f,f^\prime} = \left| \left| {y}_n(f,t) \right|^{2} - {\sigma}^{ 2}_{n,f^\prime,t} \right|$\\
        $\mathbf{K}_{n,t} = \exp\left( -\lambda\mathbf{C}_{n,t} - 1 \right)$ \\
        \For{$j=1$ \text{to} $J_\mathrm{s}$}
        {
            $\boldsymbol{\xi}_{n,t} = \left[ \bigcdot \frac{\boldsymbol{\sigma}_{n,t}^{\bigcdot 2}}{\mathbf{K}_{n,t}\left[\bigcdot \frac{|\mathbf{y}_n(:,t)|^{\bigcdot 2}}{\mathbf{K}_{n,t}^T \boldsymbol{\xi}_{n,t}} \right]^{\bigcdot \frac{\lambda\gamma}{\lambda\gamma+1}}} \right]^{\bigcdot\frac{\lambda\gamma}{\lambda\gamma+1}} $ \\
        }
        $\boldsymbol{\nu}_{n,t} = \left[\bigcdot \frac{|\mathbf{y}_n(:,t)|^{\bigcdot 2}}{\mathbf{K}_{n,t}^T \boldsymbol{\xi}_{n,t}} \right]^{\bigcdot \frac{\lambda\gamma}{\lambda\gamma+1}}$ \\
        $\hat{\mathbf{Q}}_{n,t} = \mbox{diag}(\boldsymbol{\xi}_{n,t}) \mathbf{K}_{n,t}^{\mathsf{{{T}}}} \boldsymbol{\nu}_{n,t}$ \\
        }
    }
    \BlankLine
      \SetKwInOut{Output}{\textbf{Output}}
    \Output{Optimal mapping matrix $\hat{\mathbf{Q}}_{n,t}=\mathsf{OM}\left[\boldsymbol{\sigma}^{\bigcdot 2}_{n,t},\mathbf{y}_n(:,t)\right]$.}
\end{algorithm}

\subsection{Minimization of the objective function}
\label{SectionIII-B}

In this subsection, we derive the iterative algorithm {{\cite{44}}} to solve the minimization problem in \eqref{eq:optimal_mapping}, i.e., to estimate the optimal mapping matrix $\hat{\mathbf{Q}}_{n,t}=\mathsf{OM}\left[\boldsymbol{\sigma}^{\bigcdot 2}_{n,t},\mathbf{y}_n(:,t)\right]$ based on the hypothetical source parameters and the reconstructed sources. The first-order derivative of \eqref{eq:optimal_mapping2} with respect to
$\mathbf{Q}_{n,t}$ can be written as
\begin{align}\label{SinkhornIVA11}
 & \frac{\partial \mathcal{S}_{\frac{1}{\lambda},\gamma}^2\left( \boldsymbol{\sigma}^{\bigcdot 2}_{n,t} \Big| |\mathbf{y}_n(:,t)_{\bigcdot}|^{2} \right)}{\partial \left[\mathbf{Q}_{n,t}\right]_{f,f^\prime}} \! = \! \left[\mathbf{C}_{\!n,t}\right]_{\!f\!,f^\prime} \!\! + \! \frac{1}{\lambda} \! \! \left( \! \log \left[\mathbf{Q}_{n,t}\right]_{\!f\!,f^\prime} \!\! + \! 1 \! \right) \nonumber \\
 & + \gamma  \log \left[\bigcdot \frac{\mathbf{Q}_{n,t}\mathbf{1}}{\boldsymbol{\sigma}^{\bigcdot 2}_{n,t}}\right]_{f^\prime} + \gamma  \log \left[\bigcdot \frac{\mathbf{Q}^\intercal_{n,t}\mathbf{1}}{|\mathbf{y}_n(:,t)|^{\bigcdot 2}}\right]_{f}.
\end{align}
Setting the derivative to zero yields a solution in the form of
\begin{align}\label{SinkhornIVA12-1}
\hat{\mathbf{Q}}_{n,t} = \mbox{diag}\left( \boldsymbol{\nu}_{n,t}\right)
\mathbf{K}_{n,t} \mbox{diag}\left(\boldsymbol{\xi}_{n,t}\right),
\end{align}
where \footnote{{The exponential operation is applied element-wise to the matrix.}}
\begin{align}\label{SinkhornIVA13}
 & \mathbf{K}_{n,t} = \exp\left( -\lambda\mathbf{C}_{n,t} - 1 \right),  \\
 & \boldsymbol{\nu}_{n,t} = \left[\bigcdot \frac{|\mathbf{y}_n(:,t)|^{\bigcdot 2}}{\mathbf{K}^\intercal_{n,t} \boldsymbol{\xi}_{n,t}} \right]^{\bigcdot\frac{\lambda\gamma}{\lambda\gamma+1}} , \\
 & \boldsymbol{\xi}_{n,t} = \left[\bigcdot \frac{\boldsymbol{\sigma}^{\bigcdot 2}_{n,t}}{\mathbf{K}_{n,t}\left[ \bigcdot \frac{|\mathbf{y}_n(:,t)|^{\bigcdot 2}}{\mathbf{K}^\intercal_{n,t} \boldsymbol{\xi}_{n,t}} \right]^{\bigcdot\frac{\lambda\gamma}{\lambda\gamma+1}}} \right]^{\bigcdot\frac{\lambda\gamma}{\lambda\gamma+1}}.
\end{align}
This implicit solution results in the iterative procedure outlined in Algorithm~\ref{algo:OT}.

\subsection{Sinkhorn-based IVA}

\begin{algorithm}[t]
\label{S_IVA}
\small
\setlength{\tabcolsep}{0.1mm}{
  \caption{Sinkhorn-based IVA (sIVA)}
  \SetKwInOut{Input}{\textbf{Input}}
  \Input{
         Mixture $\mathbf{x}(f,t) \in \mathbb{C}^{N}$ ($\forall f,t$),
         number of sources $N$,
         Number of iterations $J$,
         hyperparameters $\lambda, \gamma, r$.\\
        }

    \BlankLine
    \SetKwInOut{Initialization}{\textbf{Initialization}}
    \Initialization{$\boldsymbol{\sigma}_{n,t}^{\bigcdot 2}$, $\mathbf{y}(f,t)$
          $\mathbf{D}(f) = \mathbf{I}$ ($\forall f$);}
    \For{$j = 1\  \text{to}\ J$}
        {
        \For{$n = 1\  \text{to}\ N$}
            {
            \For{$t = 1\  \text{to}\ T$}
                {
                 $\hat{\mathbf{Q}}_{n,t}=\mathsf{OM}\left(\boldsymbol{\sigma}^{\bigcdot 2}_{n,t}, \mathbf{y}_n(:,t)\right)$ $\quad \textit{(see Algorith{{m}}~\ref{algo:OT})}$\\
                $\boldsymbol{{{\hat{\sigma}}}}^{\bigcdot 2}_{n,t} = \hat{\mathbf{Q}}_{n,t} \mathbf{1}$
                }
            }
%

        \For{$f = 1$ \text{to} $F$}
            {
            \For{$n = 1\  \text{to}\ N$}
                {
                $\mathbf{O}_{n,f} = \frac{1}{T} \sum_t \frac{1}{{{\hat{\sigma}}}_{n,f,t}^2} \mathbf{x}(f,t) \mathbf{x}^{{\mathsf{H}}}(f,t)$\\
                ${{\mathbf{d}}}_n(f) \leftarrow \left[\mathbf{D}(f) \mathbf{O}_{n,f} \right]^{-1} \mathbf{i}_n$ \\
                ${{\mathbf{d}}}_n(f) \leftarrow {{\mathbf{d}}}_n(f) \left[ {{\mathbf{d}}}^{{\mathsf{H}}}_n(f) \mathbf{O}_{n,f} {{\mathbf{d}}}_n(f) \right]^{-\frac{1}{2}}$\\
                }
             \For{$t = 1\  \text{to}\ T$}
                {
                $\mathbf{y}(f,t) =  \mathbf{D}(f)\mathbf{x}(f,t) $
                }
            }
        }
    }
    \SetKwInOut{Output}{\textbf{Output}}
    \Output{Separated signal ${\mathbf{y}(f,t)}$.}
\end{algorithm}

In the previous subsection, we derived the optimal estimate of the mapping matrix associated with the $n$th source
at $t$th time frame, i.e., $\hat{\mathbf{Q}}_{n,t}$, leading to an estimate of the hypothetical source parameters in \eqref{eq:optimal_mapping_sigma]}. In what follows, we discuss how this estimate can be incorporated into the traditional auxiliary IVA to enhance separation. Essentially, conventional IVA involves estimating the stationary demixing matrix  $\mathbf{D}(f)$  by iterating the following updating rules:
\begin{align}
  \label{SinkhornIVA15-2}& \mathbf{d}_n(f) \leftarrow \left[ \mathbf{D}(f) \mathbf{O}_{n,f} \right]^{-1} \mathbf{i}_n, \\
  \label{SinkhornIVA15-3}& \mathbf{d}_n(f) \leftarrow \mathbf{d}_n(f) \left[ \mathbf{d}_n^{\mathsf{H}}(f) \mathbf{O}_{n,f} \mathbf{d}_n(f) \right]^{-\frac{1}{2}},
\end{align}
where
\begin{align}
  \label{SinkhornIVA15-1}& \mathbf{O}_{n,f} = \frac{1}{T} \sum_t \frac{1}{{{\hat{\sigma}}}_{n,f,t}^2} \mathbf{x}(f,t) \mathbf{x}^{\mathsf{H}}(f,t),
\end{align}
and  $\mathbf{i}_n$ denotes the $n$th column of the identity matrix. According to the demixing process given in \eqref{SourceEst}, the sources can then be estimated as $\mathbf{y}(f,t) = \mathbf{D}(f)\mathbf{x}(f,t)$.

{{The contrast function here, is based on the alignment induced by the Sinkhorn divergence, which introduces an entropy-regularized optimal transport cost to link $p(\mathbf{y}_n(:,t)|\mathcal{D})$ and $p(\mathbf{s}_n(:,t))$. This approach modifies the original super-Gaussian assumption by adding a regularization term, which encourages the estimated distribution $p(\mathbf{y}_n(:,t)|\mathcal{D})$ to be closer to the assumed source distribution $p(\mathbf{s}_n(:,t))$ while maintaining marginal constraints.
Auxiliary variables, such as the variance parameters $\hat{\sigma}_{n,f,t}^2$ are introduced to simplify the update steps.}}
In particular, the quantity in \eqref{SinkhornIVA15-1} is driven by the hypothetical source parameters that can be updated by leveraging the optimal mapping matrix, i.e.,  ${{\boldsymbol{\hat{\sigma}}}}_{n,t}^{\bigcdot 2} = \hat{\mathbf{Q}}_{n,t} \mathbf{1}$, {{which is computed by minimizing the Sinkhorn divergence.}} Embedding this re-estimation of the source parameters into this conventional IVA framework leads to the so-called Sinkhorn-IVA (sIVA). {{These optimized variances, derived using the optimal mapping matrix, are incorporated into the demixing matrix update process. This ensures that the estimation aligns the estimated source distribution with the hypothetical source model, improving the separation performance while retaining the core structure of IVA. By embedding this regularization, the sIVA algorithm leverages optimal transport theory for more robust source separation,}} which is summarized in Algorithm~\ref{S_IVA}.

\begin{algorithm}[t!]
\label{S_ILRMA}
\small
\setlength{\tabcolsep}{0.1mm}{
  \caption{Sinkhorn-based ILRMA (sILRMA)}
  \SetKwInOut{Input}{\textbf{Input}}   
  \Input{
         Mixture $\mathbf{x}(f,t) \in \mathbb{C}^{N}$ ($\forall f,t$),
         number of sources $N$,
         Number of iterations $J$,
         hyperparameters $\lambda, \gamma, r$.\\
        }

    \BlankLine
    \SetKwInOut{Initialization}{\textbf{Initialization}}
    \Initialization{$\boldsymbol{\sigma}_{n,\cdot,t}^{\bigcdot 2}$,
          $\mathbf{D}(f) = \mathbf{I}$, ($\forall f$);}
    \For{$j = 1\  \text{to}\ J$}
        {
        \For{$n = 1\  \text{to}\ N$}
            {
            \For{$t = 1\  \text{to}\ T$}
                {
                 $\hat{\mathbf{Q}}_{n,t}=\mathsf{OM}\left(\mathbf{y}_n(:,t),\boldsymbol{\sigma}^{\bigcdot 2}_{n,t}\right)$ $\quad \textit{(see Algorithm~\ref{algo:OT})}$\\
                $\boldsymbol{{{\hat{\sigma}}}}^{\bigcdot 2}_{n,t} = \hat{\mathbf{Q}}_{n,t} \mathbf{1}$
                }
            }
        \For{$n = 1\  \text{to}\ N$}
            {
            \For{$f = 1\  \text{to}\ F$}
                {
                \For{$k = 1\  \text{to}\ K$}
                    {
                    $u_{n,f,k} \!\! \leftarrow \!\! \sqrt{\frac{\sum_t {{{\hat{\sigma}}}}_{n,f,t}^{ 2} v_{n,k,t}\left( \sum_{k^\prime} u_{n,f,k^\prime} v_{n,k^\prime,t} \right)^{-2}}{\sum_t {{{\hat{\sigma}}}}_{n,f,t}^{ 2} \left( \sum_{k^\prime} u_{n,f,k^\prime} v_{n,k^\prime,t} \right)^{-1}}}$ \\
                    $v_{n,k,t} \! \leftarrow \! \sqrt{\frac{\sum_f {{{\hat{\sigma}}}}_{n,f,t}^{ 2}  u_{n,f,k}\left( \sum_{k^\prime} u_{n,f,k^\prime} v_{n,k^\prime,t} \right)^{-2}}{\sum_f {{{\hat{\sigma}}}}_{n,f,t}^{ 2} \left( \sum_{k^\prime} u_{n,f,k^\prime} v_{n,k^\prime,t} \right)^{-1}}}$ \\
                    }
                }
            }
        \For{$n = 1\  \text{to}\ N$}
            {
            \For{$f = 1\  \text{to}\ F$}
                {
                \For{$t = 1\  \text{to}\ T$}
                    {
                    ${\sigma}_{n,f,t}^2 = \sum_{k=1}^K u_{n,f,k}v_{n,k,t},$
                    }
                }
            }
        \For{$f = 1$ \text{to} $F$}
            {
            \For{$n = 1\  \text{to}\ N$}
                {
                $\mathbf{O}_{n,f} = \frac{1}{T} \sum_t \frac{1}{\sigma_{n,f,t}^2} \mathbf{x}(f,t) \mathbf{x}^{{{\mathsf{H}}}}(f,t)$\\
                ${{\mathbf{d}}}_n(f) \leftarrow \left[\mathbf{D}(f) \mathbf{O}_{n,f} \right]^{-1} \mathbf{i}_n$ \\
                ${{\mathbf{d}}}_n(f) \leftarrow {{\mathbf{d}}}_n(f) \left[ {{\mathbf{d}}}^{{{\mathsf{H}}}}_n(f) \mathbf{O}_{n,f} {{\mathbf{d}}}_n(f) \right]^{-\frac{1}{2}}$\\
                }
             \For{$t = 1\  \text{to}\ T$}
                {
                $\mathbf{y}(f,t) =  \mathbf{D}(f)\mathbf{x}(f,t) $
                }
            }
        }
    }
    \SetKwInOut{Output}{\textbf{Output}}
    \Output{Separated signal ${\mathbf{y}(f,t)}$.}
\end{algorithm}

\subsection{Sinkhorn-based ILRMA}
ILRMA is widely recognized as a state-of-the-art algorithm for determined BSS. The conventional approach to ILRMA involves modeling the source distribution using an NMF model, i.e.,
\begin{align}\label{SinkhornILRMA1}
 p\left[{s}_n(f,t)\right] &= \frac{1}{{\sigma}_{n,f,t}^2} \exp \left( - \frac{\left| s_n(f,t) \right|^2}{{\sigma}_{n,f,t}^2} \right),
 \end{align}
 where the variance parameters are decomposed according to {{\eqref{SourceModelVariance}}}.
Approximating the MAP estimate for this model is performed using the following update rules:
\begin{align}
\label{SinkhornILRMA8-1} u_{n,f,k} \! \leftarrow \! \sqrt{\frac{\sum_t {\sigma}_{n,f,t}^{ 2} v_{n,k,t}\left( \sum_{k^\prime} u_{n,f,k^\prime} v_{n,k^\prime,t} \right)^{-2}}{\sum_t {\sigma}_{n,f,t}^{ 2} \left( \sum_{k^\prime} u_{n,f,k^\prime} v_{n,k^\prime,t} \right)^{-1}}}, \\
 \label{SinkhornILRMA8-2} v_{n,k,t} \! \leftarrow \! \sqrt{\frac{\sum_f {\sigma}_{n,f,t}^{ 2} u_{n,f,k}\left( \sum_{k^\prime} u_{n,f,k^\prime} v_{n,k^\prime,t} \right)^{-2}}{\sum_f {\sigma}_{n,f,t}^{ 2} \left( \sum_{k^\prime} u_{n,f,k^\prime} v_{n,k^\prime,t} \right)^{-1}}}.
\end{align}
Then, the demixing matrix $\mathbf{D}(f)$ of sILRMA is
iteratively updated following the same rules as IVA, i.e., {{\eqref{SinkhornIVA15-2}, \eqref{SinkhornIVA15-3}, \eqref{SinkhornIVA15-1}}}
Nevertheless, the assumption of independence across STFT bins inherent to ILRMA could potentially limit the performance of
source separation. To address this, similar to Sinkhorn IVA, the optimal mapping of the source parameters described in Section~\ref{subec:source} is  employed
to mitigate the mismatch between the estimated source distribution and
the assumed source distribution. Specifically, the optimal mapping matrix is exploited to adjust the source parameters according to $\boldsymbol{{{\hat{\sigma}}}}_{n,t}^{\bigcdot 2} = \hat{\mathbf{Q}}_{n,t} \mathbf{1}$ along the separation conducted by ILRMA. {{The so-called Sinkhorn-based ILRMA (sILRMA)'s iterative procedure is summarized in Algorithm~\ref{S_ILRMA}.}}

\begin{figure}[t]
\vspace{-0cm} 
\setlength{\abovecaptionskip}{0cm} 
\setlength{\belowcaptionskip}{-0cm} 
\centering
\subfloat[Scenario~1, Condition~1]{\includegraphics[width=1.7in]{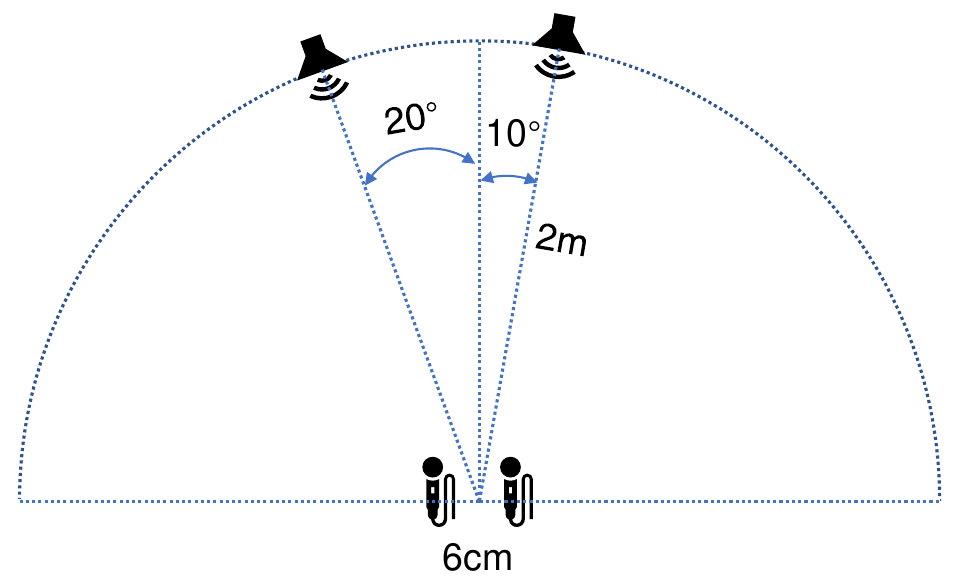}%
\label{Reuters_delta_Coh_exp}}\hspace{-3mm}
\hfil
\subfloat[Scenario~1, Condition~2]{\includegraphics[width=1.7in]{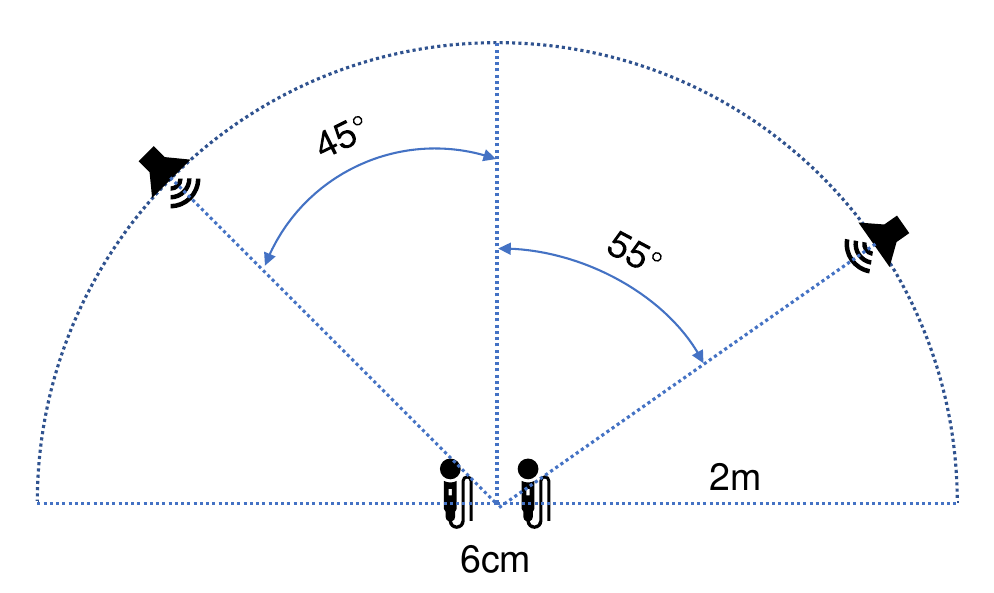}%
\label{Reuters_delta_SC_exp}}\hspace{-3mm}
\hfil
\subfloat[Scenario~2, Condition~1]{\includegraphics[width=1.7in]{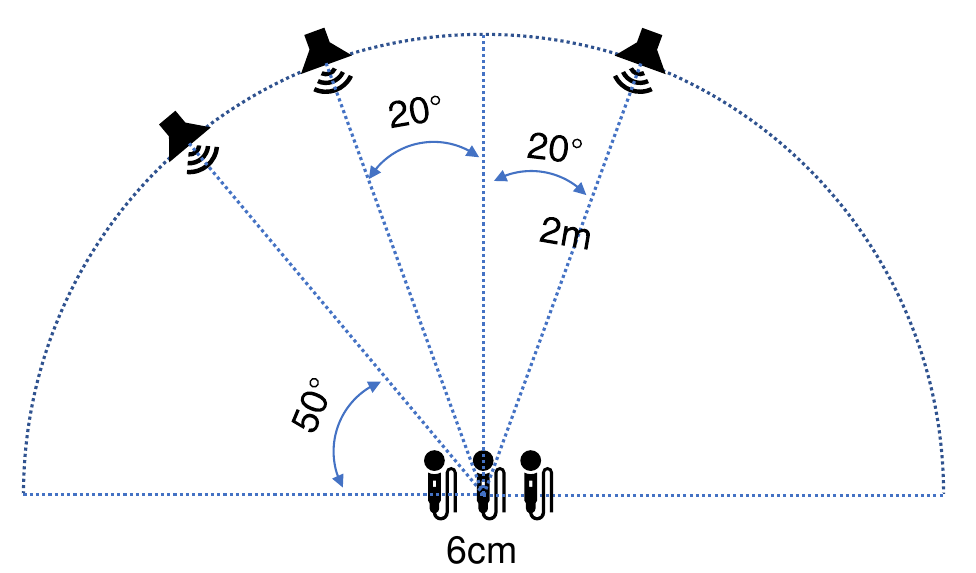}%
\label{fig18a}}\hspace{-3mm}
\hfil
\subfloat[Scenario~2, Condition~2]{\includegraphics[width=1.7in]{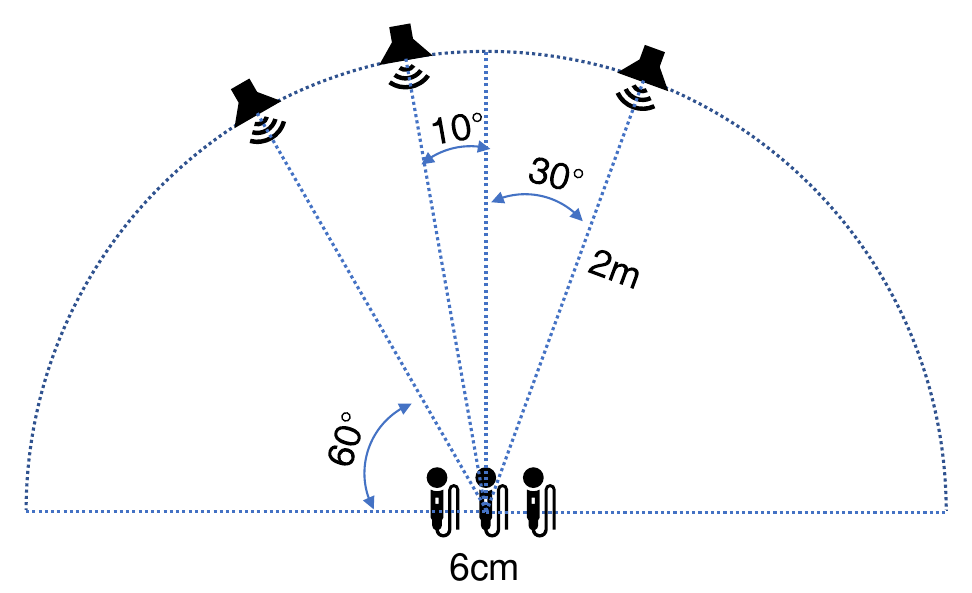}%
\label{fig18b}}\hspace{-3mm}
\hfil
\vspace{0.3cm}
\caption{ Recording conditions for the 2-channel (\fontpersosmall{Scenario~1}, 1st row) and 3-channel speech mixtures (\fontpersosmall{Scenario~2}, 2nd row).}
\vspace{-0cm}
\label{fig6}
\end{figure}

\begin{figure}[t]
\begin{tikzpicture}
\node[inner sep=0pt] (oSNR) at (-50pt,0pt)
    {\includegraphics[width=43mm]{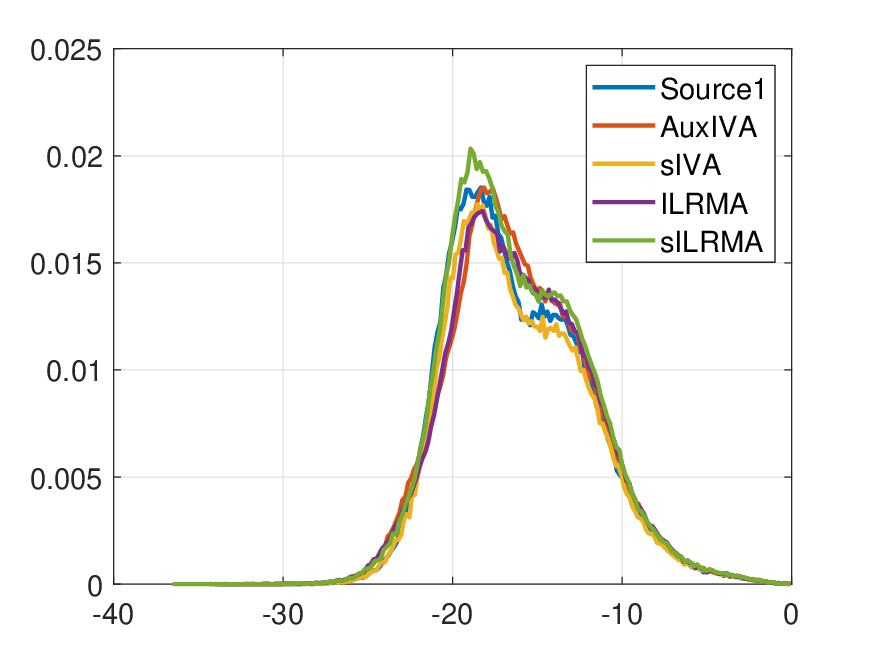}};
\draw  node[] at (5pt,-50pt) {\small The logarithm of the normalized magnitude squared spectra};
\draw  node[] at (-50pt,-65pt) {(a) Source 1};
\draw node[rotate=90,  anchor=center] at (-115pt, 0pt) {\small Probability};

\node[inner sep=0pt] (vsd) at (63pt, 0pt)
    {\includegraphics[width=43mm]{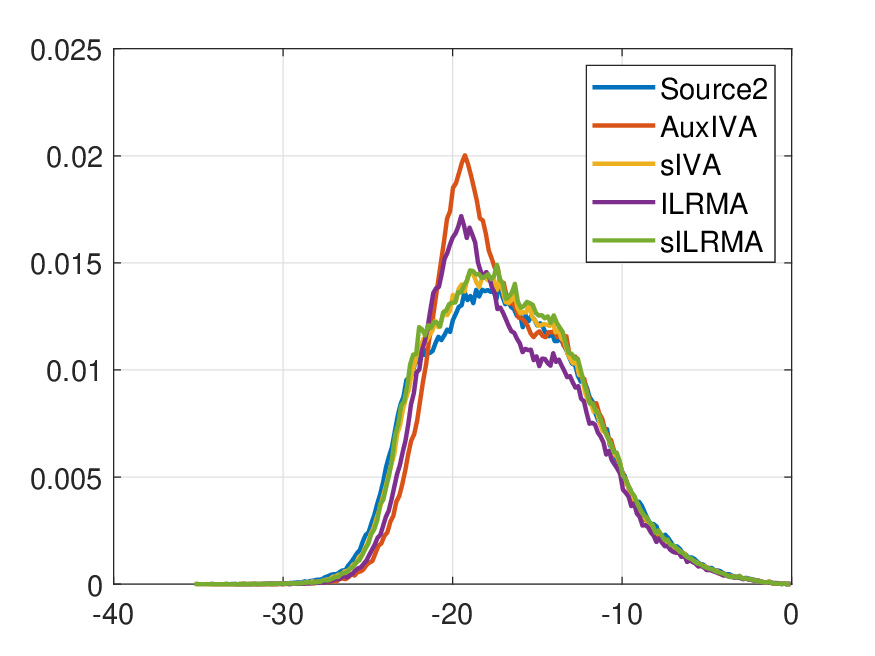}};
\draw  node[] at (60pt,-65pt) {(b) Source 2};
\end{tikzpicture}
\vspace{-0.3cm}
\caption{Fitted vs. Actual Source Distributions.}\label{figdemoplot}
\end{figure}
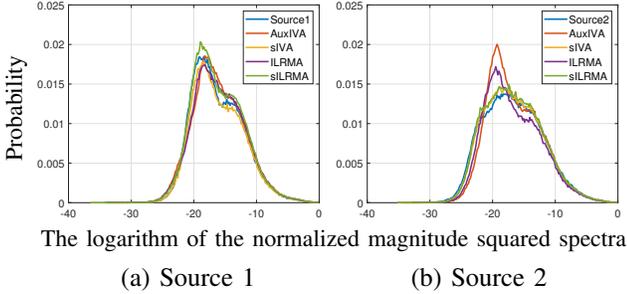

{{
\subsection{Discussion}

The computational complexity of the optimal transport (OT) algorithm, specifically when using Sinkhorn iterations, is approximately $O\left( F^2 \right)$, where $F$ denotes the number of frequency bins. This complexity arises from the iterative process of solving for the optimal transport matrix with entropy regularization, which involves matrix-vector operations to balance rows and columns and satisfy marginal constraints.
The overall complexity can increase significantly for high-resolution frequency representations or larger datasets. In such cases, strategies like optimizing the number of iterations or using approximate Sinkhorn methods can help mitigate the computational burden.

Compared to conventional IVA methods, which do not involve OT-based regularization, our Sinkhorn-based IVA method introduces additional computational overhead due to the computation of the OT matrix. However, this added complexity is justified by the enhanced separation accuracy achieved through optimal transport, as it more precisely aligns the estimated and assumed source distributions, thereby reducing model mismatch errors.
}}

\section{Experiments}
\label{Exp}

\begin{figure}[t]
\vspace{-0cm}
\begin{minipage}[b]{.46\linewidth}
  \centering
  \centerline{\includegraphics[width=4.2cm]{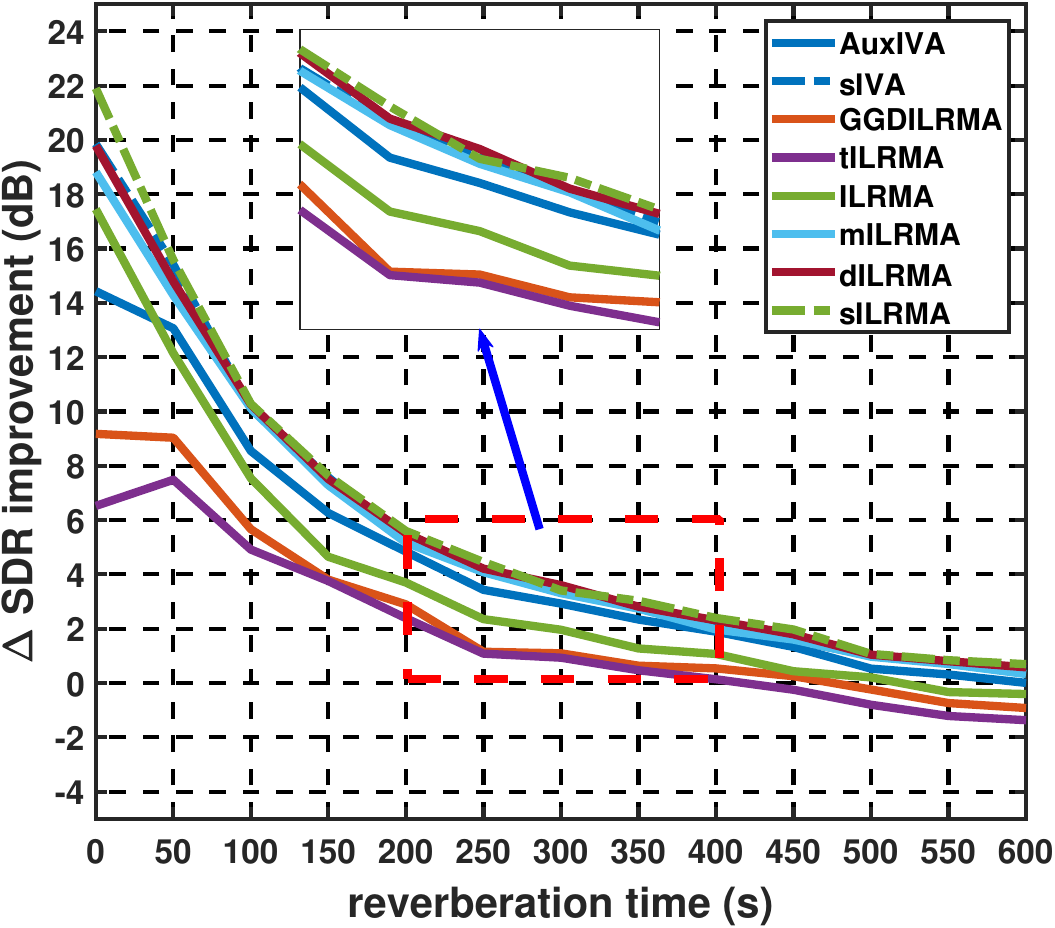}}
  \centerline{(a) female+female}\medskip
\end{minipage}
\begin{minipage}[b]{.55\linewidth}
  \centering
  \centerline{\includegraphics[width=4.2cm]{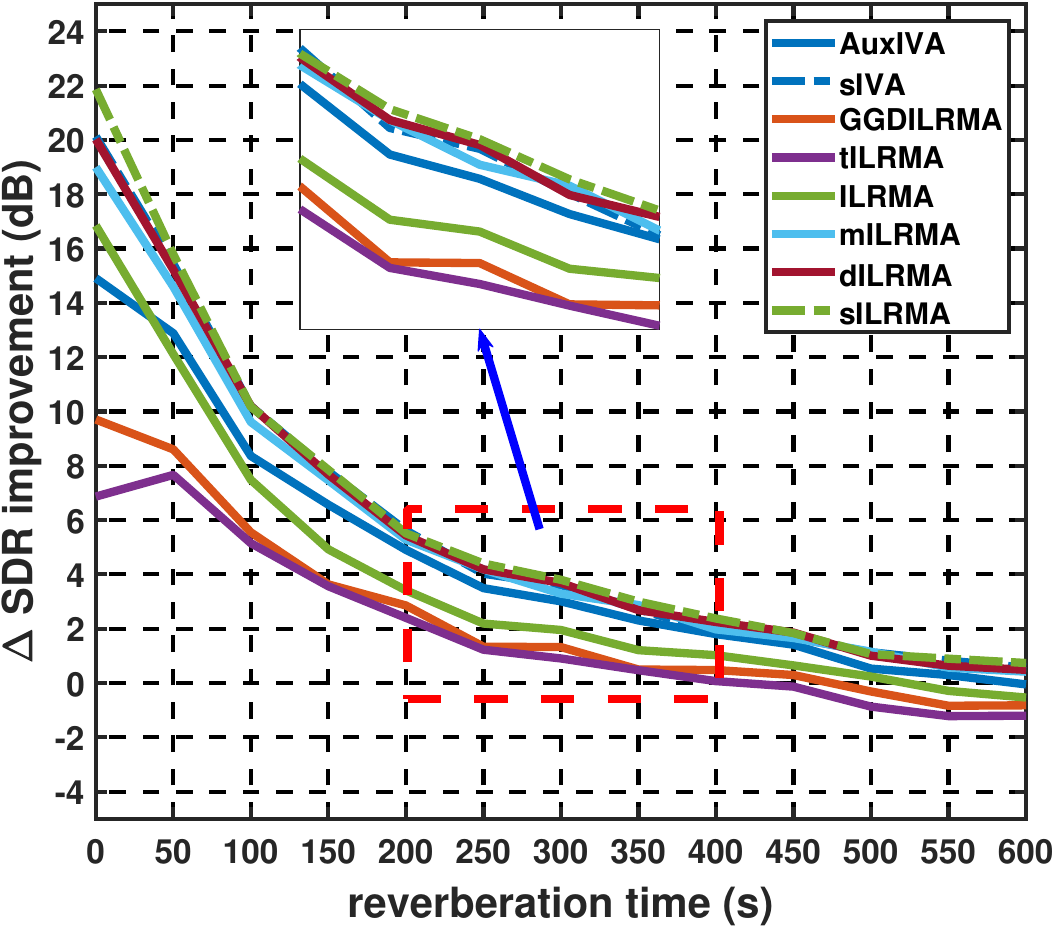}}
  \centerline{(b) female+female}\medskip
\end{minipage}
\begin{minipage}[b]{0.46\linewidth}
  \centering
  \centerline{\includegraphics[width=4.2cm]{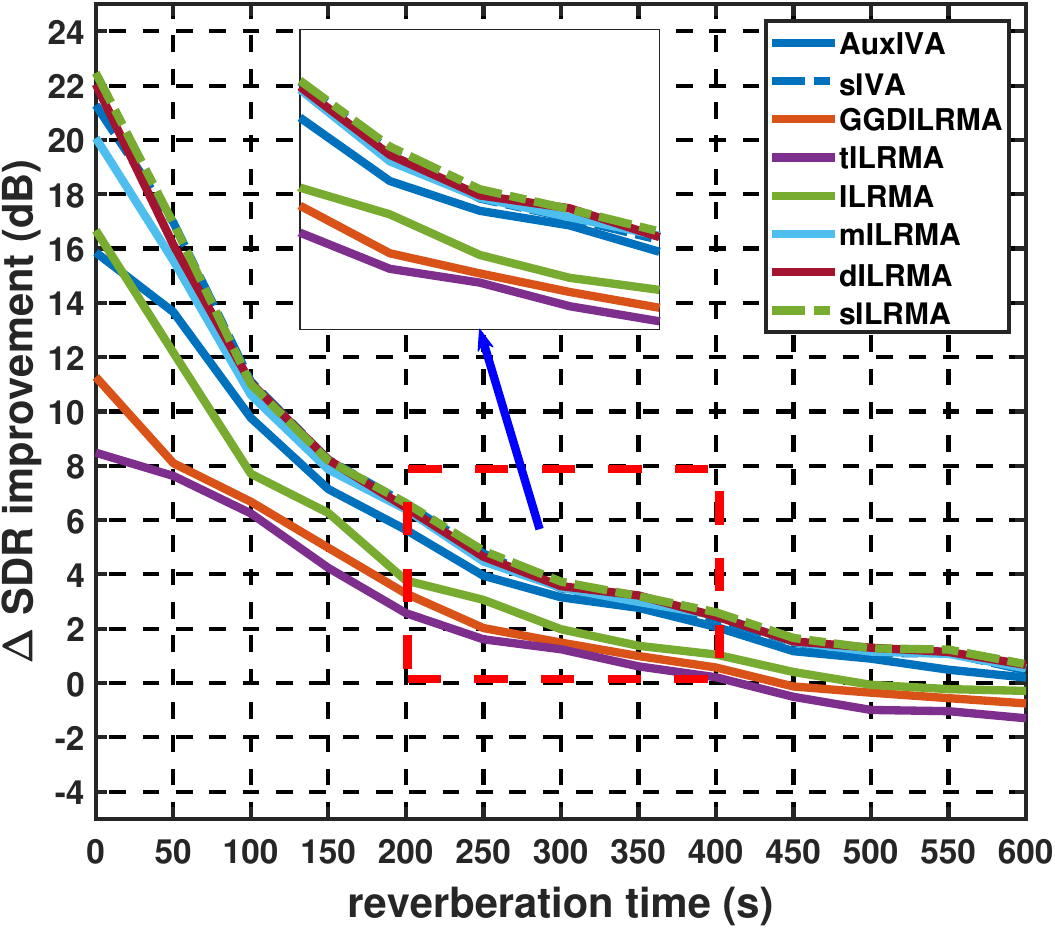}}
  \centerline{(c) male+male}\medskip
\end{minipage}
\begin{minipage}[b]{.55\linewidth}
  \centering
  \centerline{\includegraphics[width=4.2cm]{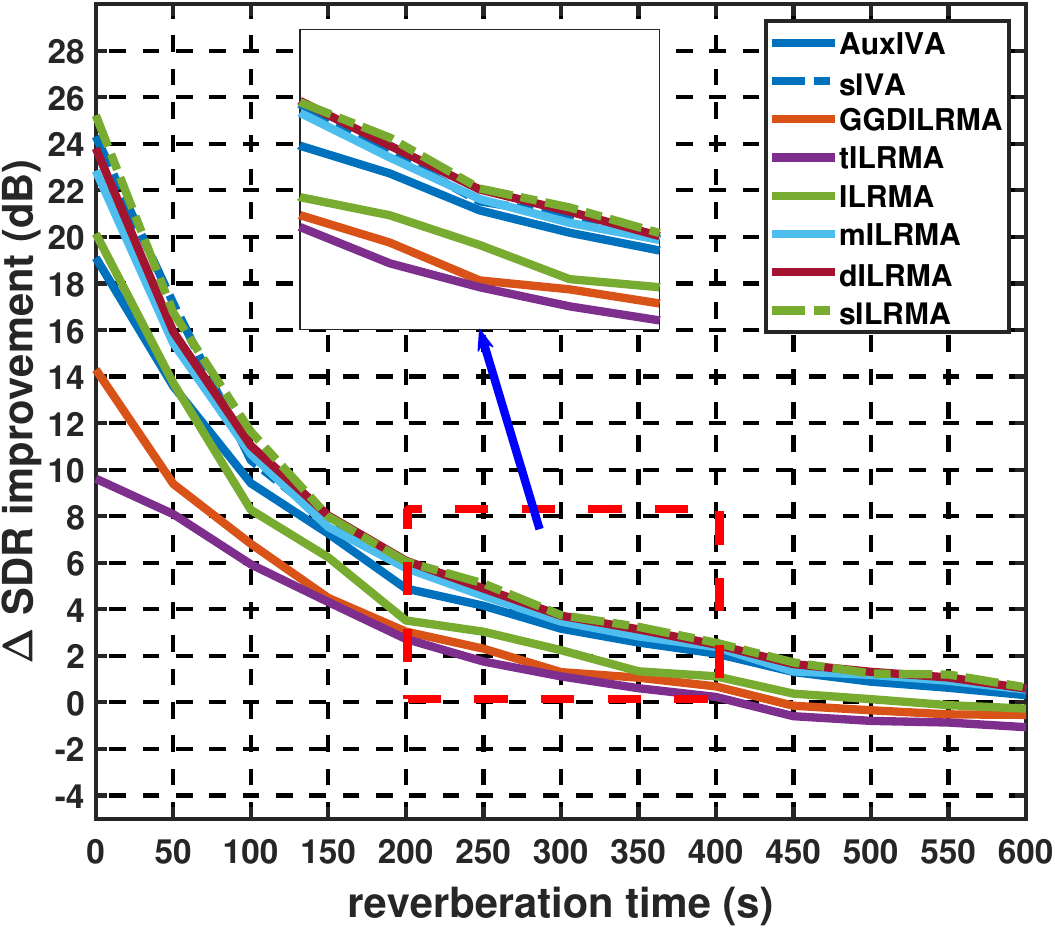}}
  \centerline{(d) male+male}\medskip
\end{minipage}
\begin{minipage}[b]{.46\linewidth}
  \centering
  \centerline{\includegraphics[width=4.2cm]{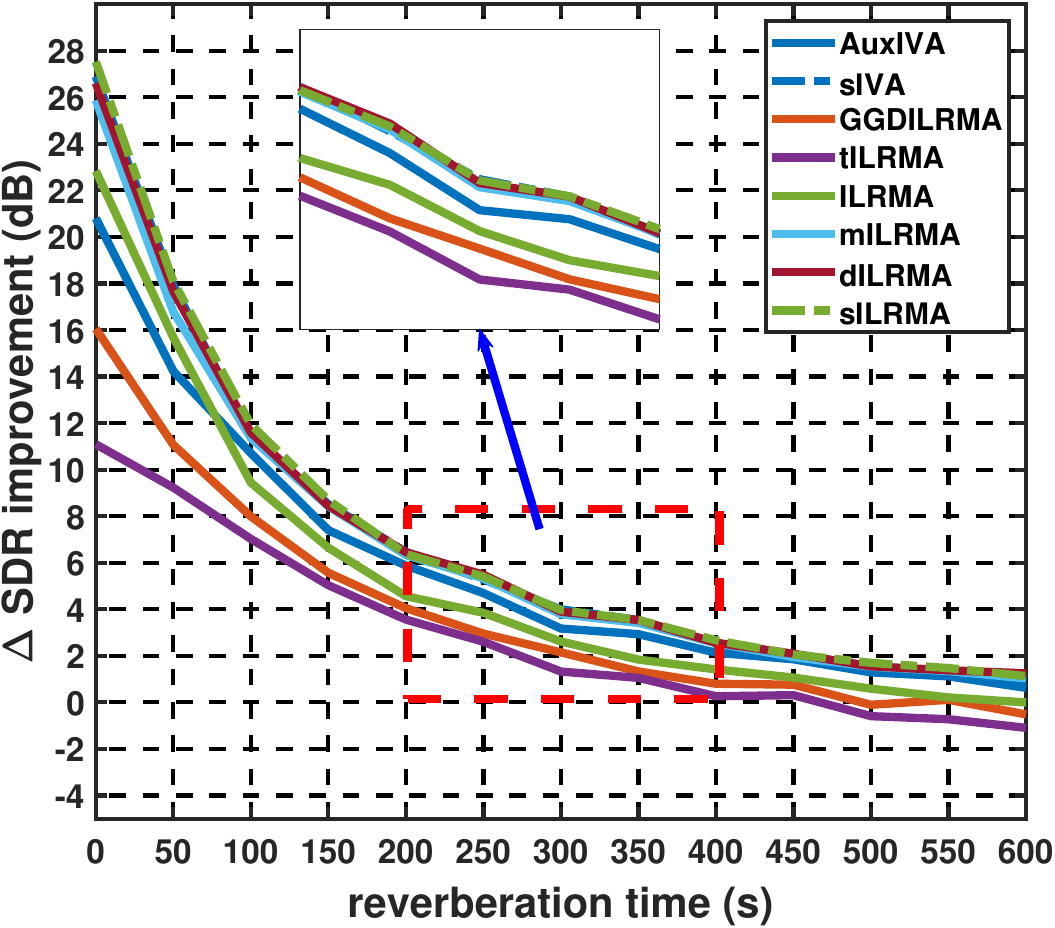}}
  \centerline{(e) female+male}\medskip
\end{minipage}
\begin{minipage}[b]{0.55\linewidth}
  \centering
  \centerline{\includegraphics[width=4.2cm]{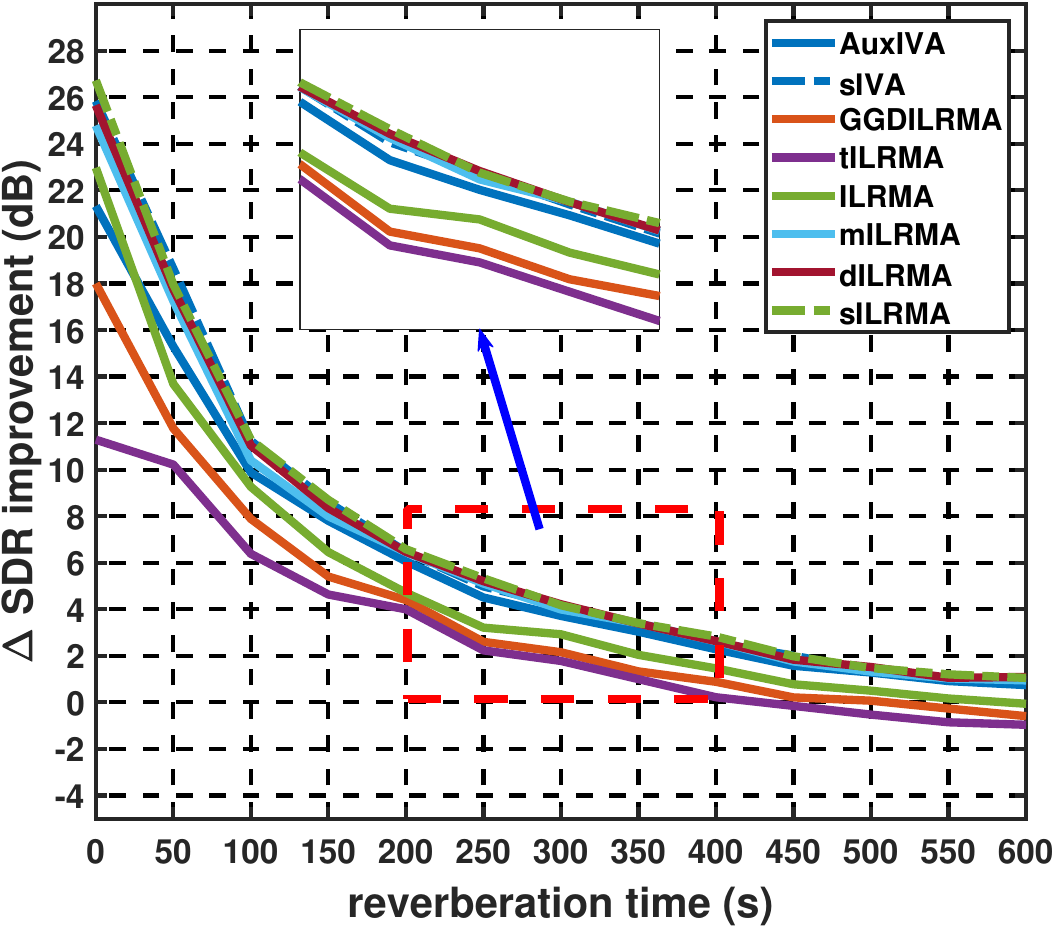}}
  \centerline{(f) female+male}\medskip
\end{minipage}
\caption{SDR improvement of the compared methods for \fontpersosmall{Scenario~1}: \fontpersosmall{Condition~1} (1st column) and \fontpersosmall{Condition~2} (2nd column).}
\label{fig8}
\vspace{-0cm}
\end{figure}

\begin{figure}[t]
\vspace{-0cm}
\begin{minipage}[b]{.46\linewidth}
  \centering
  \centerline{\includegraphics[width=4.2cm]{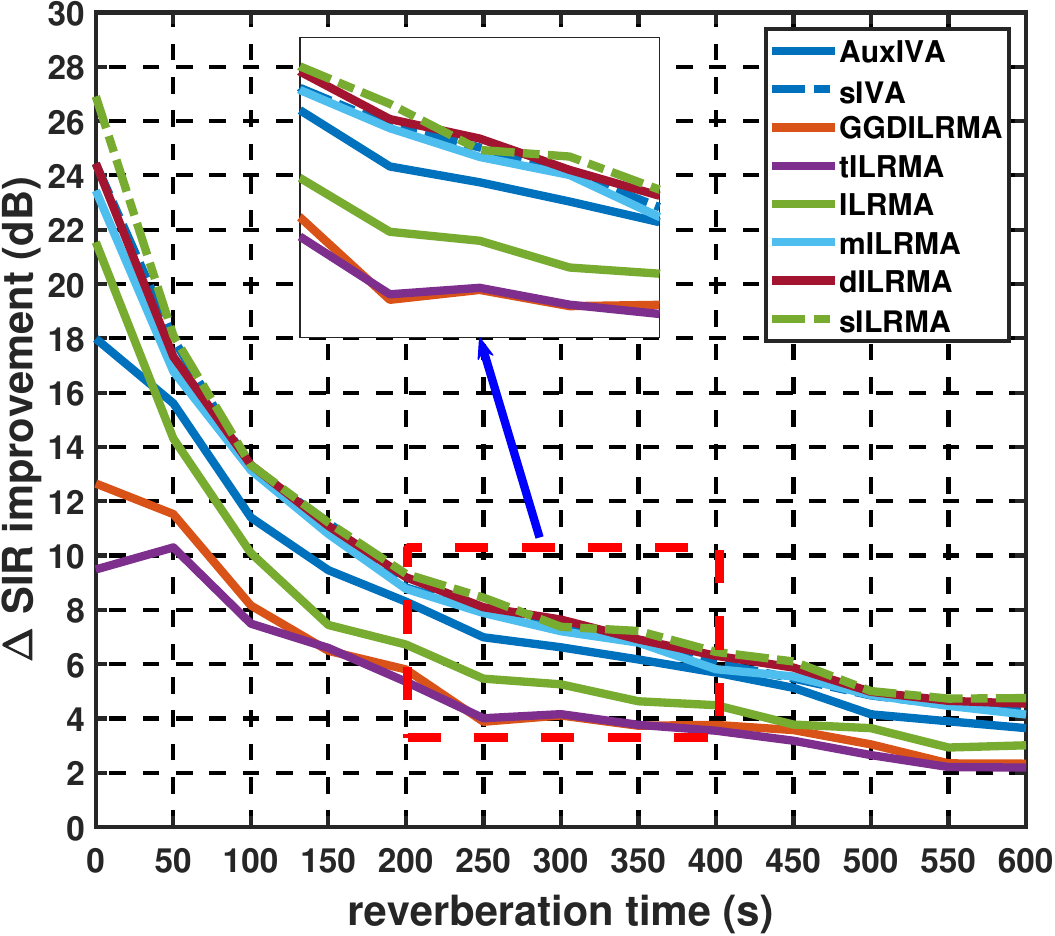}}
  \centerline{(a) female+female}\medskip
\end{minipage}
\begin{minipage}[b]{.55\linewidth}
  \centering
  \centerline{\includegraphics[width=4.2cm]{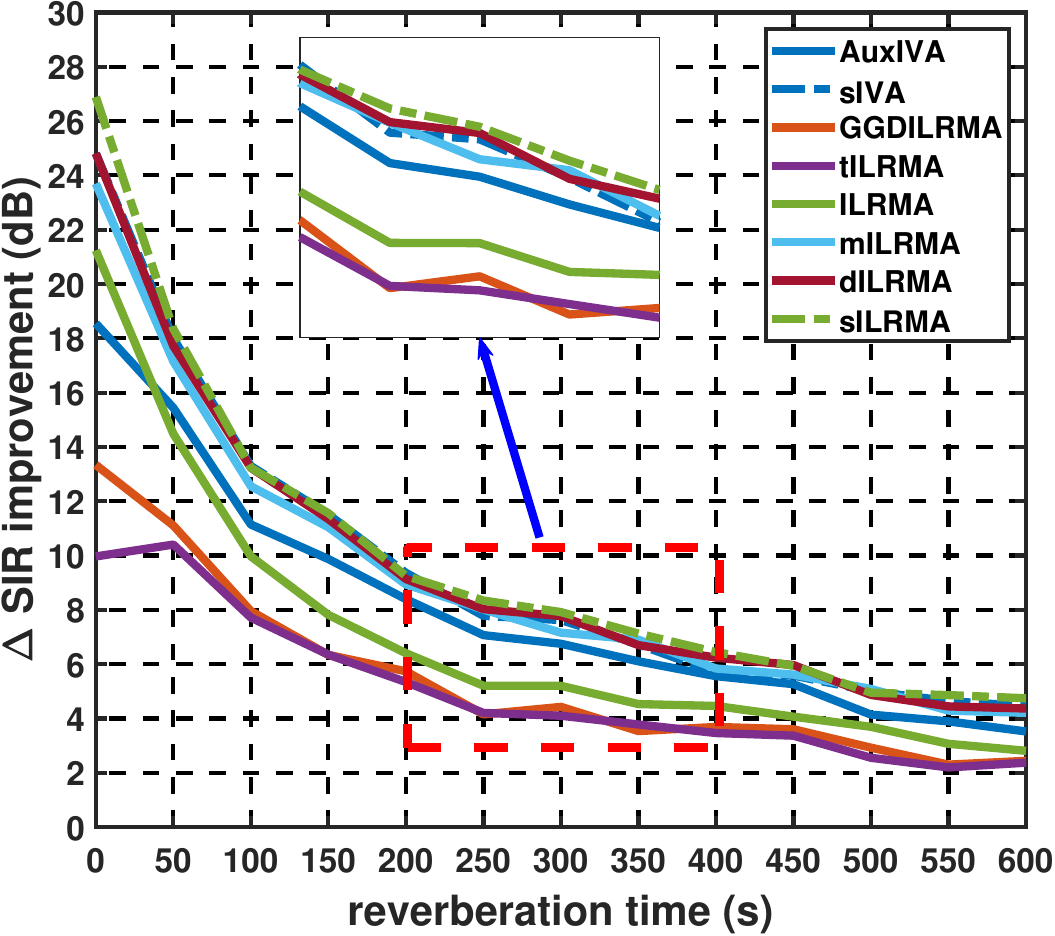}}
  \centerline{(b) female+female}\medskip
\end{minipage}
\begin{minipage}[b]{0.46\linewidth}
  \centering
  \centerline{\includegraphics[width=4.2cm]{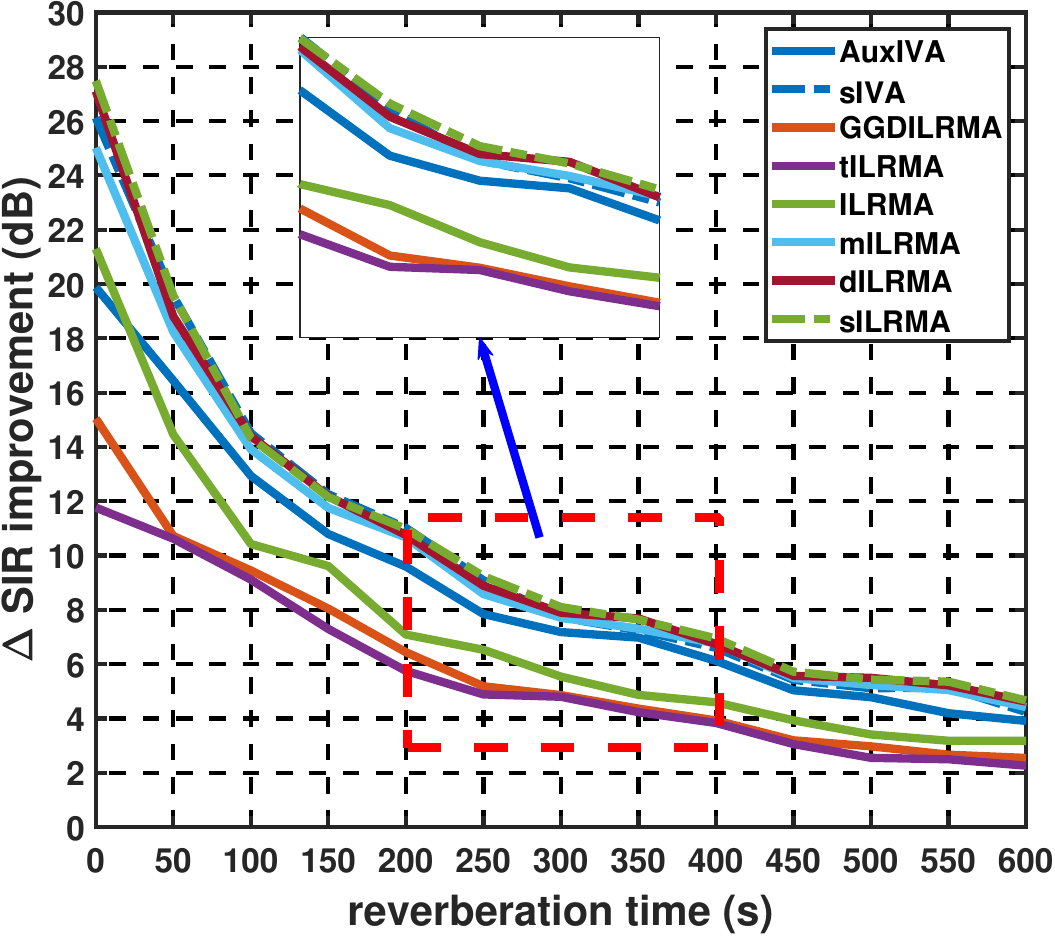}}
  \centerline{(c) male+male}\medskip
\end{minipage}
\begin{minipage}[b]{.55\linewidth}
  \centering
  \centerline{\includegraphics[width=4.2cm]{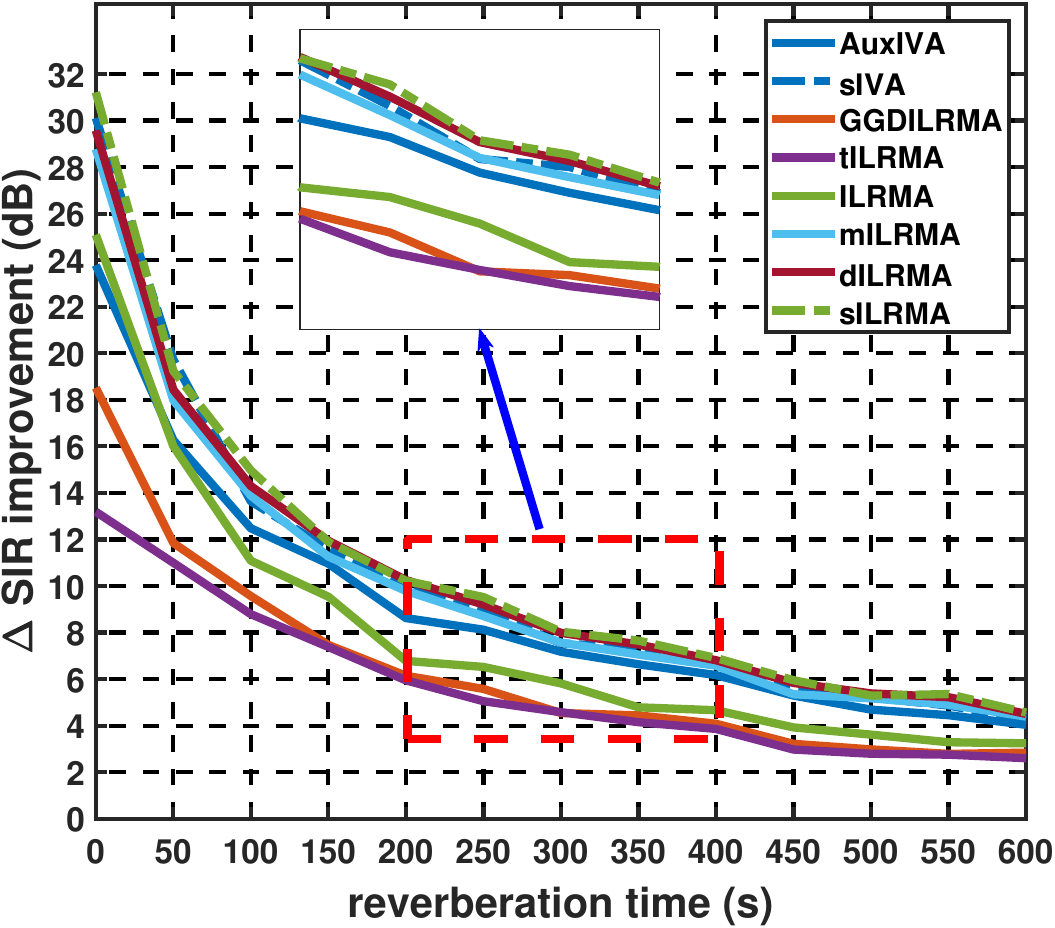}}
  \centerline{(d) male+male}\medskip
\end{minipage}
\begin{minipage}[b]{.46\linewidth}
  \centering
  \centerline{\includegraphics[width=4.2cm]{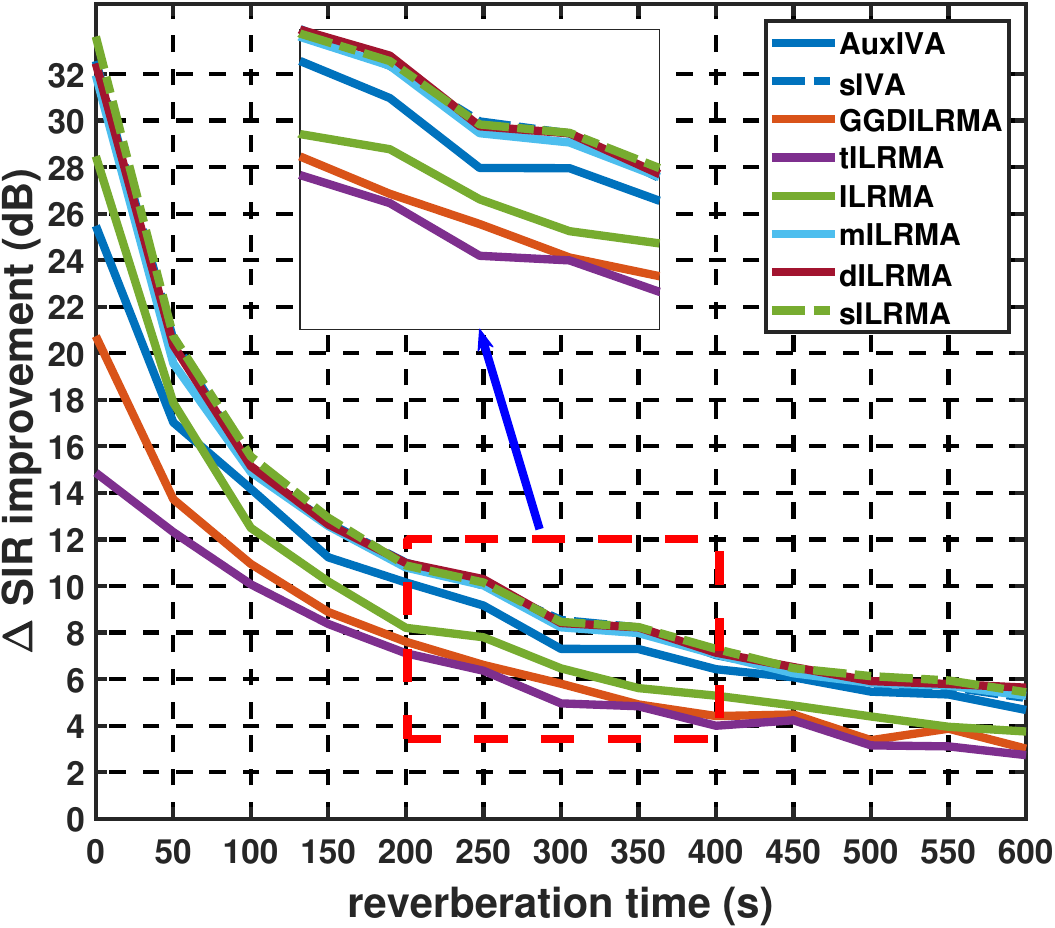}}
  \centerline{(e) female+male}\medskip
\end{minipage}
\begin{minipage}[b]{0.55\linewidth}
  \centering
  \centerline{\includegraphics[width=4.2cm]{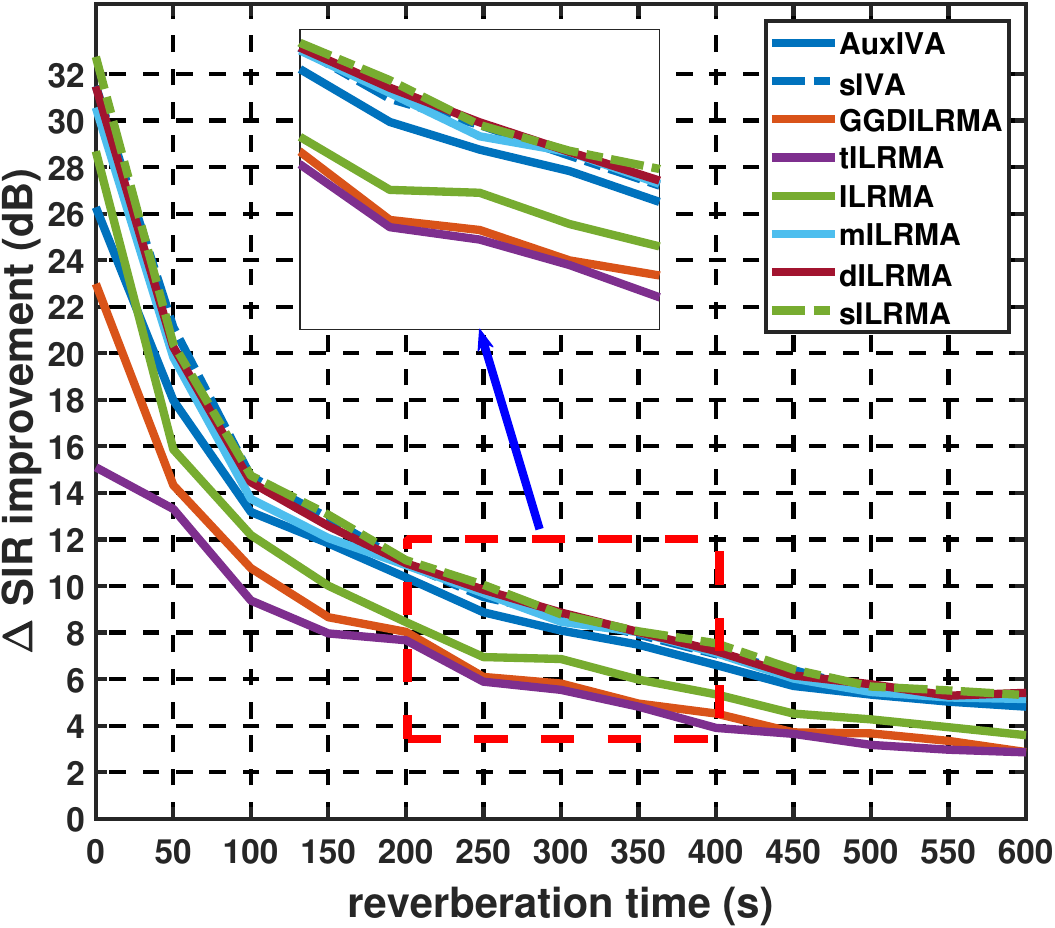}}
  \centerline{(f) female+male}\medskip
\end{minipage}
\caption{SIR improvement of the compared methods for \fontpersosmall{Scenario~1}: \fontpersosmall{Condition~1} (1st column) and \fontpersosmall{Condition~2} (2nd column).}
\label{fig10}
\vspace{-0cm}
\end{figure}

\begin{figure}[t]
\vspace{-0.cm}
\begin{minipage}[b]{.46\linewidth}
  \centering
  \centerline{\includegraphics[width=4.2cm]{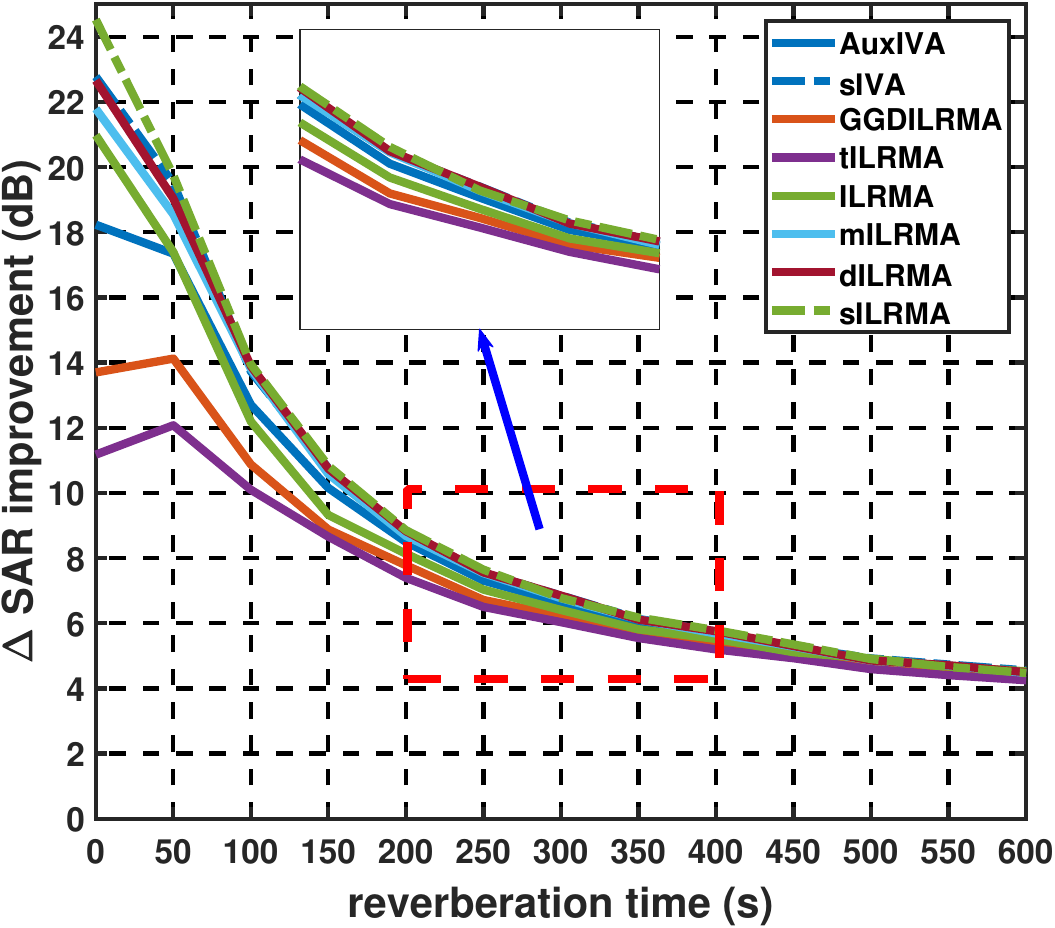}}
  \centerline{(a) female+female}\medskip
\end{minipage}
\begin{minipage}[b]{.55\linewidth}
  \centering
  \centerline{\includegraphics[width=4.2cm]{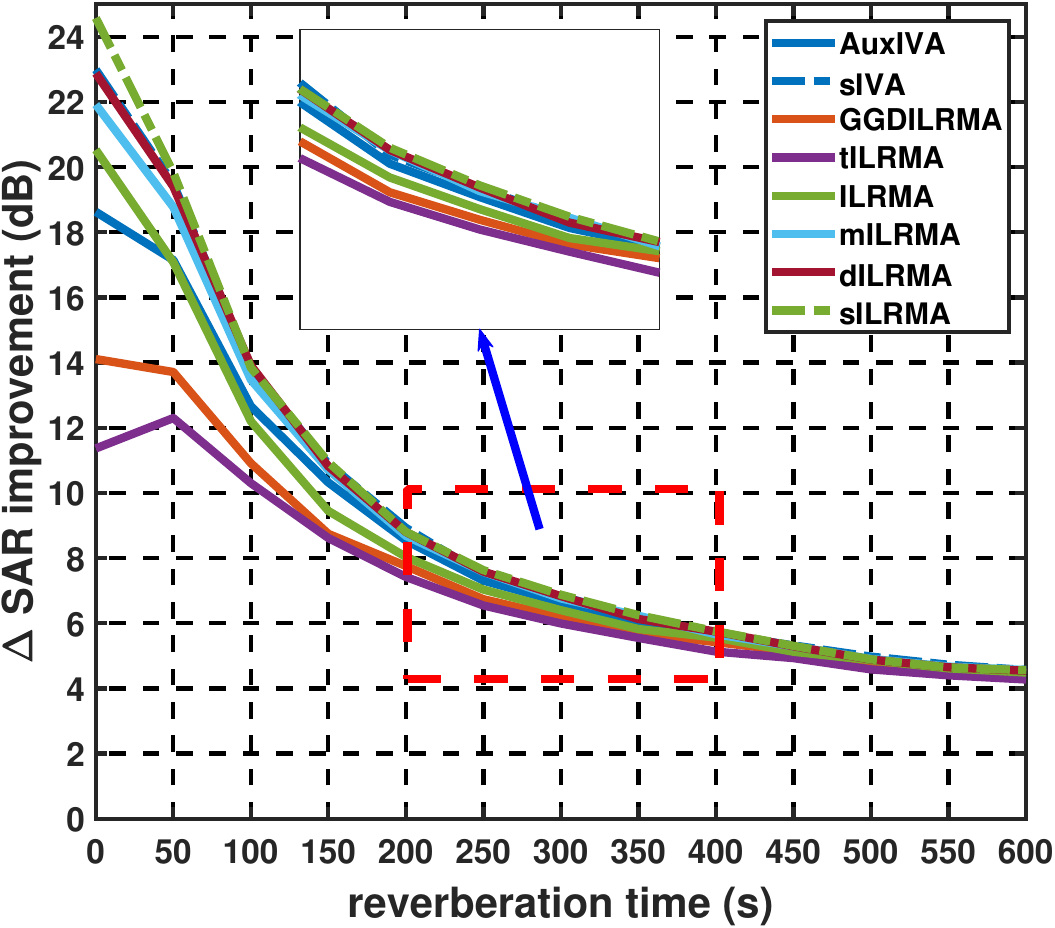}}
  \centerline{(b) female+female}\medskip
\end{minipage}
\vspace{-0.3cm}

\begin{minipage}[b]{0.46\linewidth}
  \centering
  \centerline{\includegraphics[width=4.2cm]{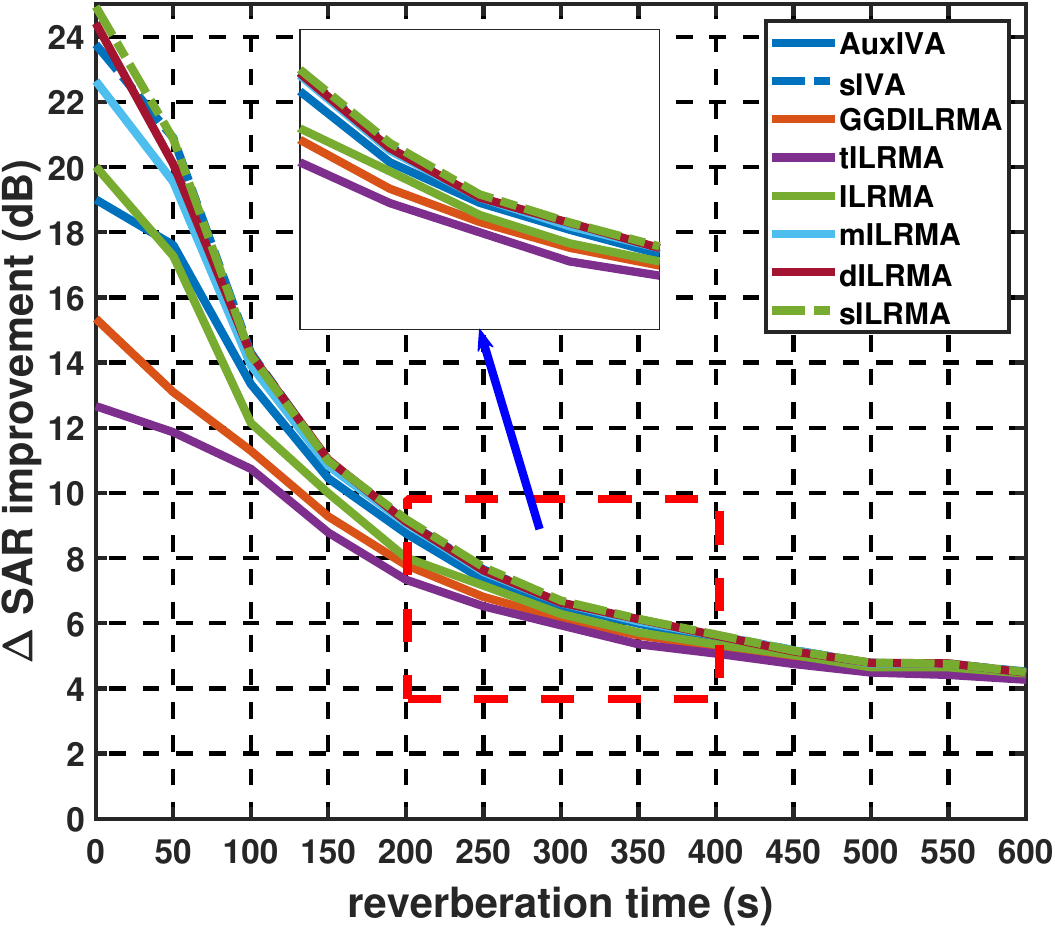}}
  \centerline{(c) male+male}\medskip
\end{minipage}
\begin{minipage}[b]{.55\linewidth}
  \centering
  \centerline{\includegraphics[width=4.2cm]{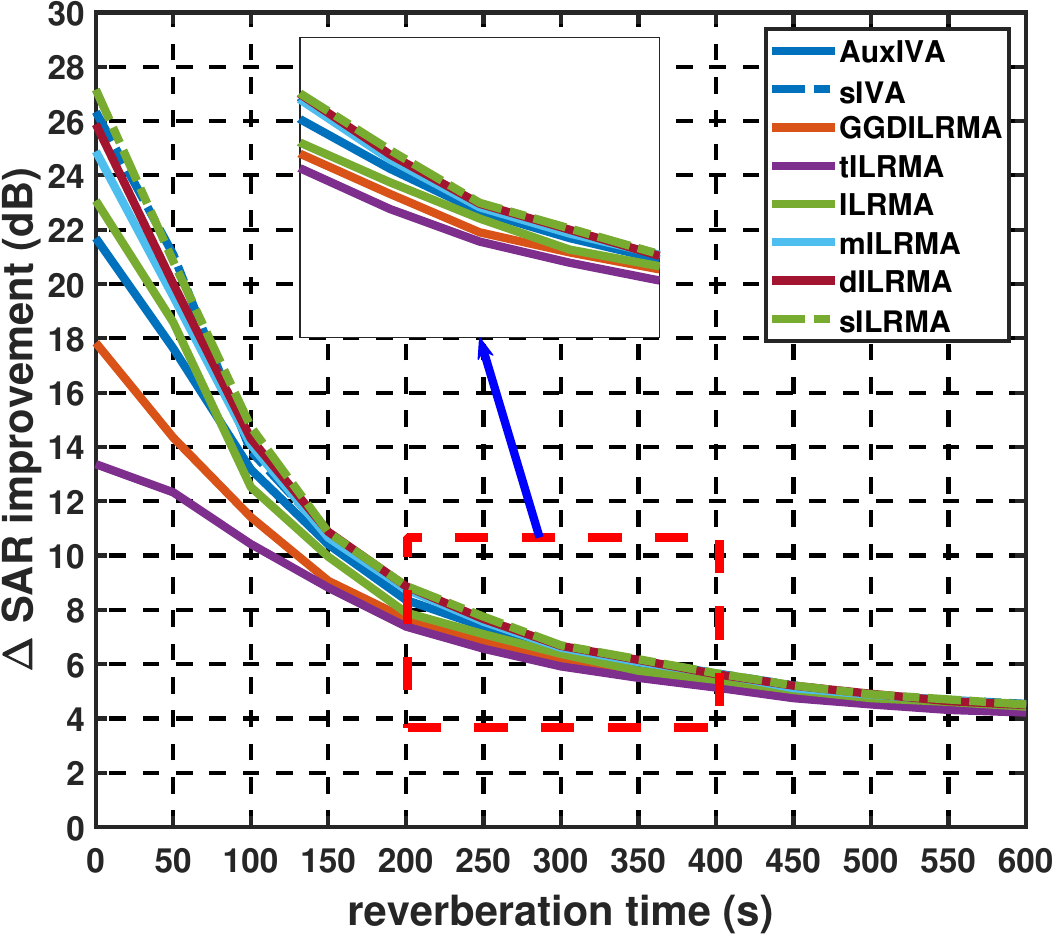}}
  \centerline{(d) male+male}\medskip
\end{minipage}
\vspace{-0.3cm}

\begin{minipage}[b]{.46\linewidth}
  \centering
  \centerline{\includegraphics[width=4.2cm]{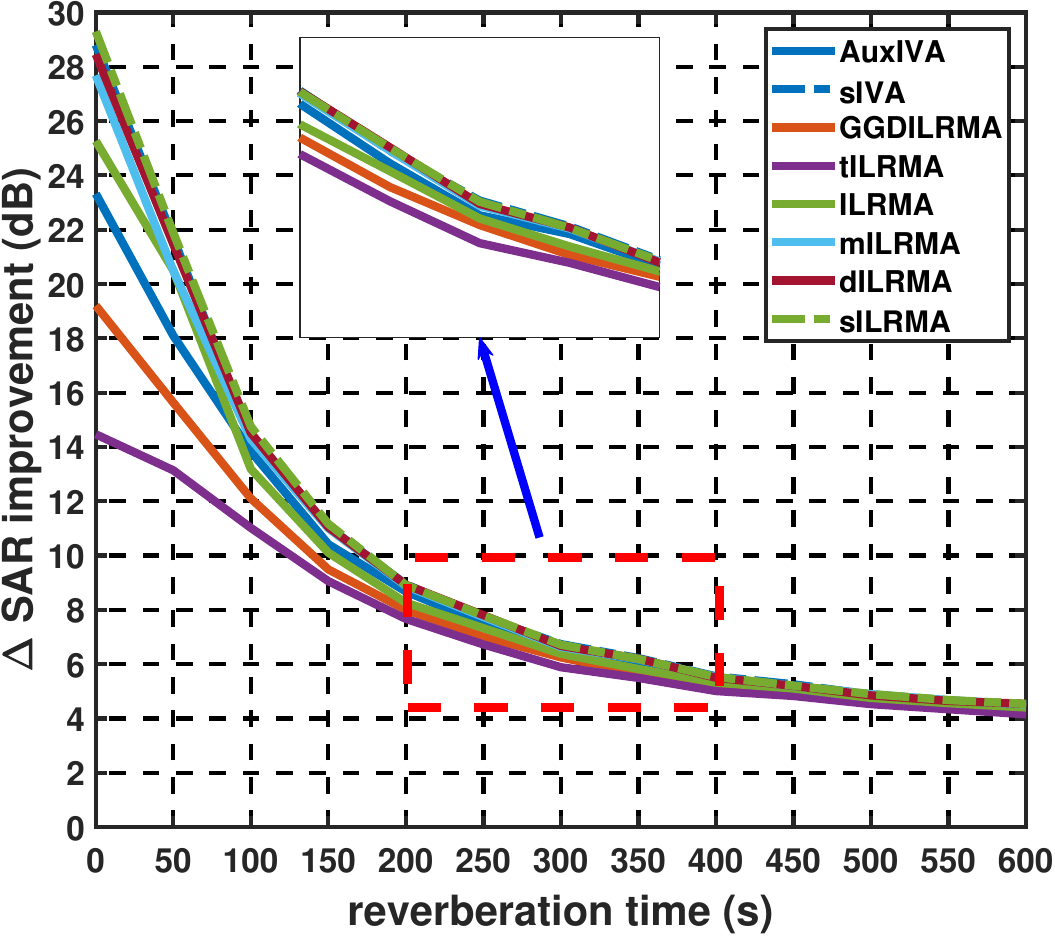}}
  \centerline{(e) female+male}\medskip
\end{minipage}
\begin{minipage}[b]{0.55\linewidth}
  \centering
  \centerline{\includegraphics[width=4.2cm]{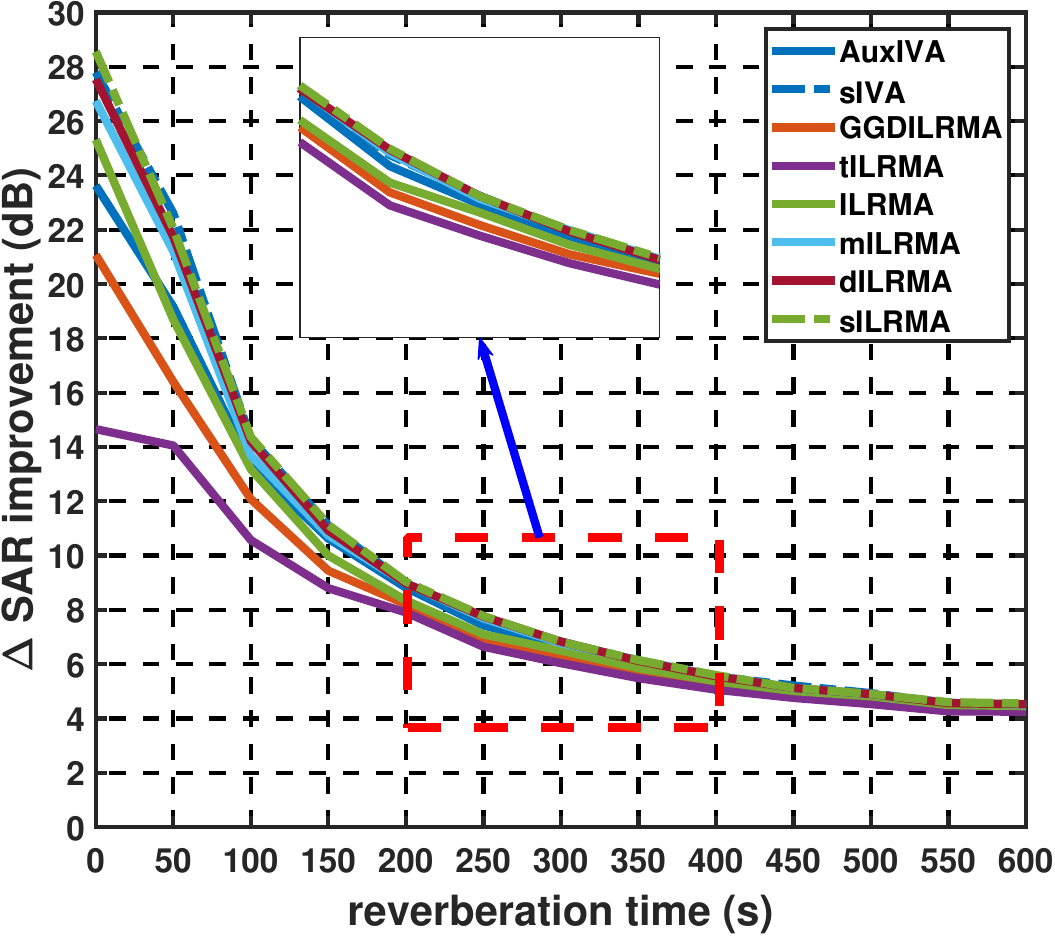}}
  \centerline{(f) female+male}\medskip
\end{minipage}
\vspace{-0.8cm}
\caption{SAR improvement of the compared methods for \fontpersosmall{Scenario~1}: \fontpersosmall{Condition~1} (1st column) and \fontpersosmall{Condition~2} (2nd column).}
\label{fig12}
\vspace{-0cm}
\end{figure}

{{The experiments are divided into three sections. First, we describe the datasets, experimental setup, comparison methods, and evaluation metrics. Next, we present histogram-based simulations to evaluate distribution alignment, followed by separation experiments with two sources and two microphones, as well as three sources and three microphones. Finally, we assess the performance of sIVA and sILRMA under different hyperparameter settings.}}


\subsection{Configuration}
\subsubsection{Simulated data sets}
The simulations utilize speech signals selected from the Wall Street Journal (WSJ0) corpus \cite{50} as clean speech sources.
Following the process outlined in the {{SiSEC}} challenge \cite{48}, simulated mixtures are generated
for specific multichannel speech separation tasks, with two configurations: $M
= N = 2$ (referred to as \fontperso{Scenario~1}) and $M = N = 3$ (referred to as
\fontperso{Scenario~2}). The image model method \cite{51} was used to generate room
impulse responses. In both cases, the room dimensions were set to $8 \times 8
\times 3$~m, with speakers positioned $2$~m away from the center
of the microphones. The reflection coefficients of all the room surfaces,
i.e., walls, ceiling and floor,  are assumed to be identical and they are
determined using Sabine's formula \cite{young1959sabine} with reverberation
time $T_{60}$ ranging from $0$~ms to $600$~ms with $50$-ms increment. Two test
conditions were considered for each case, denoted as \fontperso{Condition~1} and
\fontperso{Condition~2}. Recording settings under these two conditions for both scenarios are illustrated in Fig.~\ref{fig6}.

Three gender combinations of sources for \fontperso{Scenario~1} and four gender combinations of sources for
\fontperso{Scenario~2} are considered, exploring various reverberation conditions with different
values of $T_{60}$. The sampling rate for all cases was set at $16$~kHz. For each scenario, condition and combinations, a set 100
mixtures has been generated for evaluation.

\subsubsection{Compared algorithms and parameter settings}
The hyperparameters of sIVA and sILRMA for all experiments
have been set as follows: $\gamma = 1$, $\lambda = 5$ for sIVA and sILRMA, and $r = 2$, $K = 10$ for both
algorithms. These algorithms are compared to the following  algorithms.
\begin{itemize}
 \itemsep=0.0pt
 \item {Auxiliary-function-based IVA (AuxIVA)\cite{10}}: It achieves
     fast convergence-guaranteed optimization through alternating updates
     of auxiliary parameters and demixing matrices.

 \item {ILRMA\cite{4}}: It integrates AuxIVA and NMF to include low-rank time-frequency structures for each source.
  The iteration limit is capped at 100, with the number of basis vectors $K$ specified as $10$.

 \item {t-distribution based ILRMA (tILRMA)\cite{14}}: The source generative model in ILRMA, originally based on a complex Gaussian distribution, is now substituted with a complex Student's $t$-distribution. The hyperparameters were set to: $\nu = 1000$ and $\rho = 10$,
     respectively.

   \item {Generalized Gaussian distribution based ILRMA (GGD-ILRMA)
       \cite{33}}: It differs from tILRMA  by employing a generalized Gaussian distribution as the source generative model. The
       hyperparameters were set to:  $\beta = 1.99$ and $\rho =0.5$.

  \item {Minimum-volume ILRMA (mILRMA) \cite{16}}: It introduces
      minimum volume prior for the source model. The hyperparameters were set to: $\mu = 0.5$, $\gamma$ is initialized to $0.5$, and $K =
      10$.

  \item {{ILRMA with disjoint constraint source model (dILRMA) \cite{wang2024multichannel}}}: It
      utilizes a disjoint sparseness constraint on each source model to promote the concentration of the signal energy around the
      fundamental frequency. The hyperparameters were set to: $\rho=10$ and $\mu=0.05$.
\end{itemize}

\subsubsection{Evaluation metrics}
The source performance is assessed using the {{signal-to-distortion ratio (SDR), signal-to-interference
ratio (SIR), and the signal-to-artifacts ratio (SAR)}}, respectively defined as \cite{vincent2006performance}:
\begin{align}
\label{SDR}
\mathrm{SDR}  &\triangleq 10 \log_{10} \frac{\| s_{\mathrm{target}} \|^2}{\| e_{\mathrm{interf}} + e_{\mathrm{spat}} + e_{\mathrm{artif}} \|^2}, \\
\label{SIR}
 \mathrm{SIR} &\triangleq  10 \log_{10} \frac{\| s_{\mathrm{target}} + e_{\mathrm{spat}} \|^2}{\| e_{\mathrm{interf}} \|^2},\\
\label{SAR}
 \mathrm{SAR} &\triangleq  10 \log_{10} \frac{\| s_{\mathrm{target}} + e_{\mathrm{interf}} + e_{\mathrm{spat}} \|^2}{\| e_{\mathrm{artif}} \|^2},
\end{align}
where $s_{\mathrm{target}}$ denotes the true target source signal, and
$e_{\mathrm{spat}}$, $e_{\mathrm{interf}}$, and $e_{\mathrm{artif}}$ are the error components representing spatial distortion,
interference, and artifacts, respectively. For a more comprehensive definition of these metrics, please refer to \cite{vincent2006performance}.

\begin{table}[t]
\centering
\setlength{\abovecaptionskip}{0cm}
\caption{Average $\Delta$SDR, $\Delta$SIR and $\Delta$SAR for \fontpersosmall{Condition 1}.}\label{tab1}
\scriptsize
\setlength{\tabcolsep}{0.5mm}{
\resizebox{.99\columnwidth}{!}{
\begin{tabular}{cccccccccc}
  \toprule
    & \#RT(ms) &AuxIVA&sIVA&ILRMA&GDDILRMA&tILRMA&mILRMA&dILRMA&sILRMA \\
   \toprule
   \multirow{13}{*}{$\Delta$SDR}
          & 0     & 12.04  & 15.39  & 11.86  & 11.18  & 8.77  & 8.86  & 15.45  & 16.43  \\
          & 50    & 11.28  & 13.66  & 11.48  & 10.42  & 8.24  & 8.51  & 13.71  & 14.06  \\
          & 100   & 8.83  & 9.76  & 8.17  & 7.48  & 6.27  & 6.42  & 9.72  & 9.97  \\
          & 150   & 6.17  & 6.57  & 5.91  & 5.41  & 4.35  & 4.79  & 7.02  & 7.14  \\
          & 200   & 4.64  & 5.10  & 4.61  & 3.97  & 3.22  & 3.61  & 5.24  & 5.30  \\
          & 250   & 3.49  & 3.73  & 3.29  & 2.89  & 2.05  & 2.67  & 4.00  & 4.00  \\
          & 300   & 2.51  & 2.76  & 2.31  & 2.04  & 1.36  & 1.96  & 3.09  & 3.14  \\
          & 350   & 1.85  & 1.98  & 1.57  & 1.36  & 0.56  & 1.20  & 2.29  & 2.30  \\
          & 400   & 1.48  & 1.70  & 1.16  & 0.82  & 0.22  & 0.78  & 1.74  & 1.79  \\
          & 450   & 0.95  & 0.96  & 0.57  & 0.31  & -0.35  & 0.42  & 1.22  & 1.34  \\
          & 500   & 0.48  & 0.66  & 0.26  & -0.02  & -0.59  & 0.11  & 0.74  & 0.82  \\
          & 550   & 0.25  & 0.25  & 0.08  & -0.20  & -0.71  & 0.00  & 0.51  & 0.62  \\
          & 600   & -0.03  & 0.09  & -0.28  & -0.51  & -1.09  & -0.36  & 0.26  & 0.23  \\

  \toprule
   {{\textbf{rank (Rev)}}} &  50 $\sim$ 600       & \textbf{4.00}  & \textbf{2.92}  & \textbf{4.92}  & \textbf{6.33}  & \textbf{8.00}  & \textbf{6.67}  & \textbf{1.92}  & \textbf{1.08}  \\
   rank (All) &  0 $\sim$ 600        & 4.00  & 2.92  & 4.92  & 6.31  & 8.00  & 6.69  & 1.92  & 1.08  \\

  \toprule

  \multirow{9}{*}{$\Delta$SIR}
          & 0     & 15.53  & 19.77  & 15.76  & 14.68  & 11.97  & 12.29  & 19.79  & 21.08  \\
          & 50    & 14.18  & 16.75  & 14.32  & 13.17  & 11.16  & 11.57  & 16.80  & 17.20  \\
          & 100   & 11.66  & 12.76  & 10.76  & 10.06  & 9.01  & 9.19  & 12.67  & 12.94  \\
          & 150   & 9.21  & 9.74  & 8.78  & 8.25  & 7.23  & 7.72  & 10.28  & 10.41  \\
          & 200   & 7.90  & 8.50  & 7.77  & 7.04  & 6.34  & 6.70  & 8.70  & 8.76  \\
          & 250   & 6.89  & 7.21  & 6.58  & 6.14  & 5.35  & 5.82  & 7.61  & 7.59  \\
          & 300   & 5.95  & 6.27  & 5.63  & 5.37  & 4.79  & 5.21  & 6.75  & 6.82  \\
          & 350   & 5.36  & 5.52  & 4.91  & 4.73  & 4.01  & 4.46  & 5.94  & 5.95  \\
          & 400   & 5.08  & 5.35  & 4.58  & 4.22  & 3.79  & 4.08  & 5.42  & 5.48  \\
          & 450   & 4.51  & 4.51  & 4.01  & 3.77  & 3.28  & 3.77  & 4.93  & 5.07  \\
          & 500   & 4.04  & 4.25  & 3.70  & 3.44  & 3.09  & 3.49  & 4.43  & 4.51  \\
          & 550   & 3.88  & 3.85  & 3.59  & 3.33  & 3.06  & 3.42  & 4.26  & 4.38  \\
          & 600   & 3.61  & 3.73  & 3.23  & 3.06  & 2.76  & 3.08  & 4.06  & 4.01  \\

  \toprule
  {{\textbf{rank (Rev)}}} &  50 $\sim$ 600       & \textbf{3.92}  & \textbf{3.00}  & \textbf{4.92}  & \textbf{6.25}  & \textbf{8.00}  & \textbf{6.67}  & \textbf{1.92}  & \textbf{1.17}  \\
  rank (All) &  0 $\sim$ 600        & 4.00  & 3.00  & 4.85  & 6.23  & 8.00  & 6.69  & 1.92  & 1.15  \\

  \toprule

  \multirow{9}{*}{$\Delta$SAR}
          & 0     & 12.79  & 15.41  & 12.63  & 12.24  & 9.94  & 9.77  & 15.52  & 16.24  \\
          & 50    & 12.39  & 14.34  & 12.77  & 11.96  & 9.74  & 9.71  & 14.33  & 14.64  \\
          & 100   & 10.08  & 10.66  & 10.08  & 9.54  & 8.26  & 8.22  & 10.73  & 10.93  \\
          & 150   & 7.61  & 7.80  & 7.81  & 7.49  & 6.56  & 6.69  & 8.02  & 8.09  \\
          & 200   & 6.12  & 6.34  & 6.37  & 6.06  & 5.42  & 5.63  & 6.37  & 6.42  \\
          & 250   & 5.07  & 5.18  & 5.25  & 5.06  & 4.44  & 4.85  & 5.23  & 5.26  \\
          & 300   & 4.34  & 4.43  & 4.49  & 4.34  & 3.81  & 4.25  & 4.49  & 4.51  \\
          & 350   & 3.80  & 3.87  & 3.94  & 3.84  & 3.29  & 3.74  & 3.92  & 3.94  \\
          & 400   & 3.42  & 3.51  & 3.55  & 3.41  & 2.95  & 3.40  & 3.48  & 3.51  \\
          & 450   & 3.09  & 3.13  & 3.17  & 3.05  & 2.58  & 3.13  & 3.10  & 3.14  \\
          & 500   & 2.82  & 2.90  & 2.96  & 2.83  & 2.37  & 2.90  & 2.84  & 2.86  \\
          & 550   & 2.58  & 2.62  & 2.74  & 2.62  & 2.16  & 2.75  & 2.59  & 2.64  \\
          & 600   & 2.41  & 2.47  & 2.53  & 2.39  & 1.92  & 2.54  & 2.38  & 2.41  \\

  \toprule
  {{\textbf{rank (Rev)}}} &  50 $\sim$ 600        & \textbf{5.33}  & \textbf{3.25}  & \textbf{2.08}  & \textbf{5.75}  & \textbf{7.83}  & \textbf{5.42}  & \textbf{3.67}  & \textbf{1.83}  \\
  rank (All) &  0 $\sim$ 600        & 5.23  & 3.23  & 2.31  & 5.77  & 7.77  & 5.62  & 3.54  & 1.77  \\

  \toprule

\end{tabular}}
}
\end{table}

\subsection{Performance results}
\subsubsection{Distribution Analysis}
We utilize two 31-second clean speech signals\footnote{\tiny \textsf{book\_00000\_chp\_0009\_reader\_06709\_0.wav} and \textsf{book\_00002\_chp\_0005\_reader\_11980\_0.wav}} from the DNS dataset \cite{dubey2023icassp}. These signals are partitioned into overlapping frames with a frame size of 512 points and a $50\%$ overlap. Each frame is weighted with a Hamming window and then transformed into the STFT domain using a 1024-point fast Fourier transform (FFT). The magnitude squared spectra are subsequently computed and normalized by their maximum value. The logarithm of the normalized magnitude squared spectra is taken, and the source distributions are determined based on the histograms of these normalized magnitude squared spectra. These two source signals are then mixed according to the setup Scenario 1 with a reverberation time $T_{60} = 50$ ms. Different algorithms are applied for source separation, generating the estimated source signals. Similarly, the distributions of the mixed and estimated source signals are determined from the histograms of their respective normalized magnitude squared spectra.
Figure~\ref{figdemoplot} shows the distributions of the source, mixed, and estimated source signals for $513$ frequency bins. It is evident from Fig.~\ref{figdemoplot} that the distributions of the estimated sources using sIVA and sILRMA better match the distributions of the original source signals compared to AuxIVA and ILRMA. These results demonstrate that the Sinkhorn divergence can improve signal separation with better distribution recovery.

\begin{table}[t]
\centering
\setlength{\abovecaptionskip}{0cm}
\caption{Average $\Delta$SDR, $\Delta$SIR and $\Delta$SAR for \fontpersosmall{Condition 2}.}
\label{tab2}
\scriptsize
\setlength{\tabcolsep}{0.5mm}{
\resizebox{.99\columnwidth}{!}{
\begin{tabular}{cccccccccc}
  \toprule
    & \#RT(ms) &AuxIVA&sIVA&ILRMA&GDDILRMA&tILRMA&mILRMA&dILRMA&sILRMA \\
   \toprule
   \multirow{13}{*}{$\Delta$SDR}
          & 0     & 11.42  & 16.02  & 11.51  & 10.62  & 8.55  & 9.01  & 15.30  & 16.05  \\
          & 50    & 10.91  & 13.63  & 11.14  & 10.39  & 8.53  & 8.61  & 13.83  & 14.56  \\
          & 100   & 8.10  & 9.14  & 8.25  & 7.43  & 6.25  & 6.47  & 9.71  & 9.88  \\
          & 150   & 6.22  & 6.85  & 5.90  & 5.39  & 4.46  & 4.89  & 7.00  & 7.17  \\
          & 200   & 4.57  & 5.01  & 4.40  & 3.97  & 3.11  & 3.47  & 5.23  & 5.28  \\
          & 250   & 3.39  & 3.69  & 3.32  & 2.88  & 2.32  & 2.63  & 4.14  & 4.15  \\
          & 300   & 2.56  & 2.82  & 2.31  & 1.96  & 1.31  & 1.87  & 3.05  & 3.11  \\
          & 350   & 1.94  & 2.22  & 1.67  & 1.27  & 0.66  & 1.21  & 2.27  & 2.28  \\
          & 400   & 1.44  & 1.56  & 1.06  & 0.80  & 0.21  & 0.94  & 1.64  & 1.74  \\
          & 450   & 1.03  & 1.18  & 0.65  & 0.34  & -0.26  & 0.48  & 1.23  & 1.22  \\
          & 500   & 0.58  & 0.78  & 0.36  & 0.05  & -0.57  & 0.27  & 0.93  & 0.95  \\
          & 550   & 0.23  & 0.52  & -0.02  & -0.22  & -0.78  & -0.05  & 0.58  & 0.64  \\
          & 600   & -0.04  & 0.09  & -0.30  & -0.52  & -1.12  & -0.27  & 0.13  & 0.20  \\

  \toprule
  {{\textbf{rank (Rev)}}} &  50 $\sim$ 600        & \textbf{4.17}  & \textbf{3.00}  & \textbf{4.92}  & \textbf{6.42}  & \textbf{8.00}  & \textbf{6.50}  & \textbf{1.83}  & \textbf{1.08}  \\
   rank (All) &    0 $\sim$ 600      & 4.25  & 2.83  & 4.83  & 6.33  & 8.00  & 6.58  & 2.00  & 1.08  \\
  \toprule

  \multirow{9}{*}{$\Delta$SIR}
          & 0     & 14.76  & 20.45  & 15.38  & 14.09  & 11.74  & 12.50  & 19.74  & 20.67  \\
          & 50    & 13.70  & 16.60  & 13.91  & 13.07  & 11.43  & 11.68  & 16.85  & 17.63  \\
          & 100   & 10.79  & 12.02  & 10.83  & 9.96  & 8.96  & 9.23  & 12.66  & 12.82  \\
          & 150   & 9.25  & 10.07  & 8.78  & 8.23  & 7.38  & 7.81  & 10.24  & 10.47  \\
          & 200   & 7.84  & 8.40  & 7.52  & 7.07  & 6.22  & 6.53  & 8.70  & 8.76  \\
          & 250   & 6.74  & 7.16  & 6.59  & 6.12  & 5.65  & 5.82  & 7.77  & 7.79  \\
          & 300   & 6.01  & 6.34  & 5.62  & 5.26  & 4.72  & 5.07  & 6.69  & 6.76  \\
          & 350   & 5.45  & 5.83  & 5.04  & 4.62  & 4.16  & 4.45  & 5.92  & 5.92  \\
          & 400   & 5.01  & 5.14  & 4.47  & 4.21  & 3.79  & 4.30  & 5.29  & 5.41  \\
          & 450   & 4.62  & 4.81  & 4.09  & 3.77  & 3.39  & 3.85  & 4.92  & 4.90  \\
          & 500   & 4.19  & 4.43  & 3.83  & 3.52  & 3.13  & 3.68  & 4.67  & 4.68  \\
          & 550   & 3.86  & 4.23  & 3.43  & 3.29  & 3.01  & 3.39  & 4.33  & 4.40  \\
          & 600   & 3.60  & 3.73  & 3.18  & 3.01  & 2.69  & 3.21  & 3.85  & 3.94  \\

  \toprule
  {{\textbf{rank (Rev)}}} &  50 $\sim$ 600        & \textbf{4.17}  & \textbf{3.00}  & \textbf{4.92}  & \textbf{6.42}  & \textbf{8.00}  & \textbf{6.50}  & \textbf{1.83}  & \textbf{1.08}  \\
  rank (All) &    0 $\sim$ 600      & 4.23  & 2.92  & 4.85  & 6.38  & 8.00  & 6.54  & 1.92  & 1.08  \\

  \toprule

  \multirow{9}{*}{$\Delta$SAR}
          & 0     & 12.42  & 15.93  & 12.31  & 11.76  & 9.73  & 9.87  & 15.33  & 15.94  \\
          & 50    & 12.19  & 14.43  & 12.60  & 12.05  & 9.94  & 9.78  & 14.54  & 15.14  \\
          & 100   & 9.70  & 10.37  & 10.10  & 9.51  & 8.22  & 8.27  & 10.71  & 10.85  \\
          & 150   & 7.64  & 7.95  & 7.77  & 7.47  & 6.60  & 6.77  & 8.05  & 8.14  \\
          & 200   & 6.08  & 6.26  & 6.29  & 6.04  & 5.37  & 5.55  & 6.33  & 6.39  \\
          & 250   & 5.08  & 5.20  & 5.27  & 5.03  & 4.54  & 4.78  & 5.31  & 5.33  \\
          & 300   & 4.37  & 4.46  & 4.52  & 4.34  & 3.81  & 4.26  & 4.51  & 4.52  \\
          & 350   & 3.85  & 3.93  & 3.97  & 3.79  & 3.30  & 3.77  & 3.91  & 3.92  \\
          & 400   & 3.45  & 3.51  & 3.53  & 3.40  & 2.93  & 3.43  & 3.47  & 3.49  \\
          & 450   & 3.12  & 3.18  & 3.21  & 3.07  & 2.58  & 3.14  & 3.13  & 3.14  \\
          & 500   & 2.81  & 2.90  & 2.97  & 2.84  & 2.33  & 2.92  & 2.88  & 2.90  \\
          & 550   & 2.58  & 2.64  & 2.76  & 2.62  & 2.14  & 2.72  & 2.64  & 2.66  \\
          & 600   & 2.38  & 2.46  & 2.56  & 2.43  & 1.95  & 2.54  & 2.41  & 2.43  \\

  \toprule
  {{\textbf{rank (Rev)}}} &  50 $\sim$ 600        & \textbf{5.58}  & \textbf{3.08}  & \textbf{2.08}  & \textbf{6.00}  & \textbf{7.92}  & \textbf{5.42}  & \textbf{3.42}  & \textbf{2.08}  \\
  rank (All) &   0 $\sim$ 600       & 5.46  & 3.00  & 2.31  & 6.23  & 7.85  & 5.54  & 3.31  & 2.00  \\
  \toprule

\end{tabular}}
}
\end{table}


\subsubsection{\fontperso{Scenario~1}}
The performances of the compared algorithms are evaluated on the WSJ0 dataset with different combinations of speaker genders and under two recording conditions. The frame length is again set to 512 with a $50\%$ overlap, and the FFT size is 1024. Figures~\ref{fig8}, \ref{fig10}, and \ref{fig12} present the average SDR, SIR, and SAR improvements. In the anechoic environment, the proposed sIVA and sILRMA algorithms demonstrate significantly better performance compared to the other algorithms studied. They achieve an SDR improvement approximately 5 dB higher than AuxIVA and ILRMA, and 3 dB higher than mILRMA and dILRMA.

In the studied reverberant environments, sIVA achieves an SDR improvement of 1 to 2 dB compared to AuxIVA, while sILRMA shows an SDR improvement of more than 2 dB over ILRMA. Additionally, sILRMA's performance improvement remains superior to that of the highly competitive dILRMA algorithm. Similar trends are observed for the other evaluation metrics, namely SIR and SAR.


\begin{figure}[t]
\vspace{-0cm} 
\setlength{\abovecaptionskip}{0cm} 
\setlength{\belowcaptionskip}{-0cm} 
\centering
\subfloat[dev1]{\includegraphics[width=3.4in]{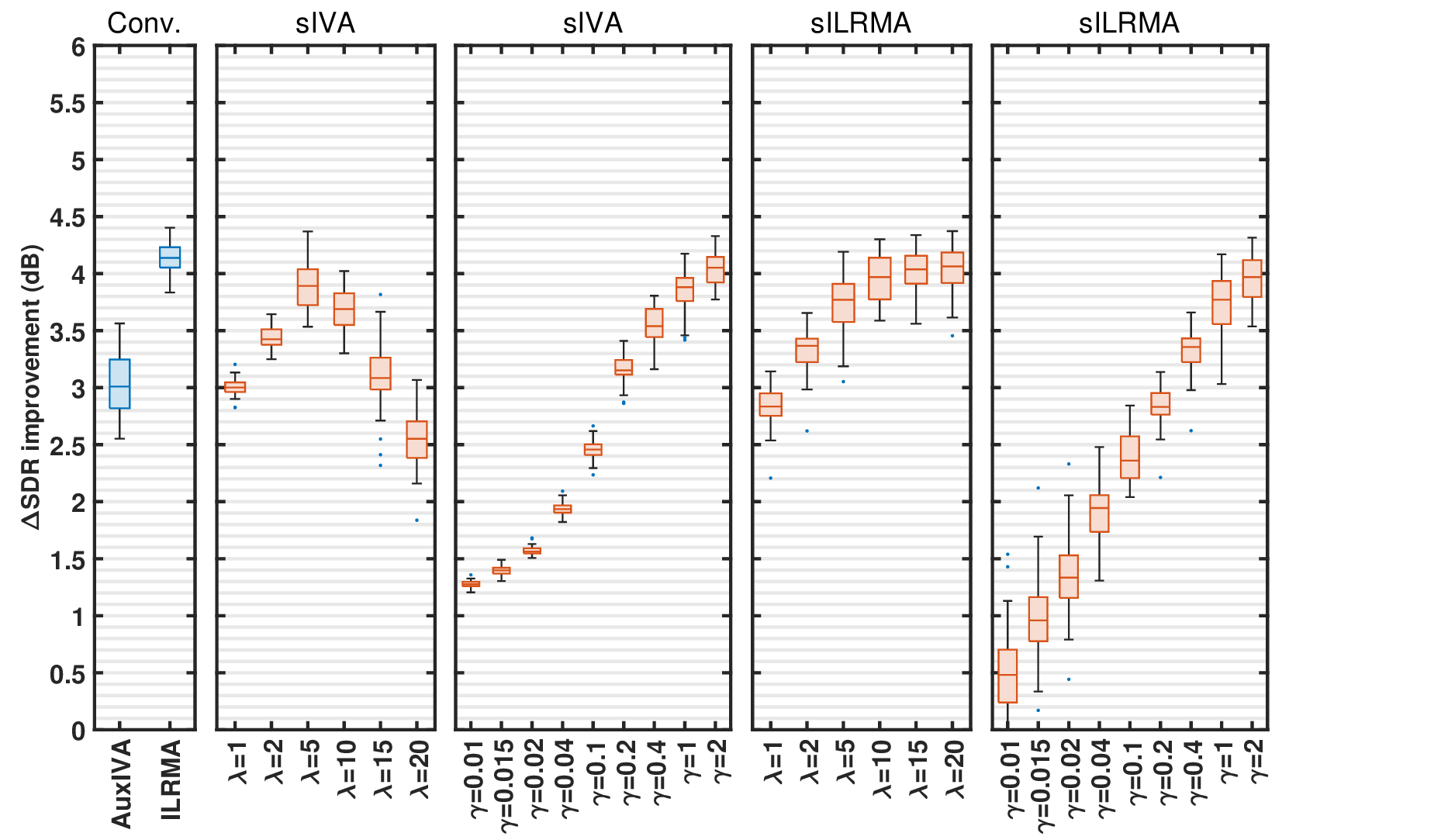}%
\label{Reuters_delta_Coh}}\hspace{-1mm}
\hfil
\subfloat[dev2]{\includegraphics[width=3.4in]{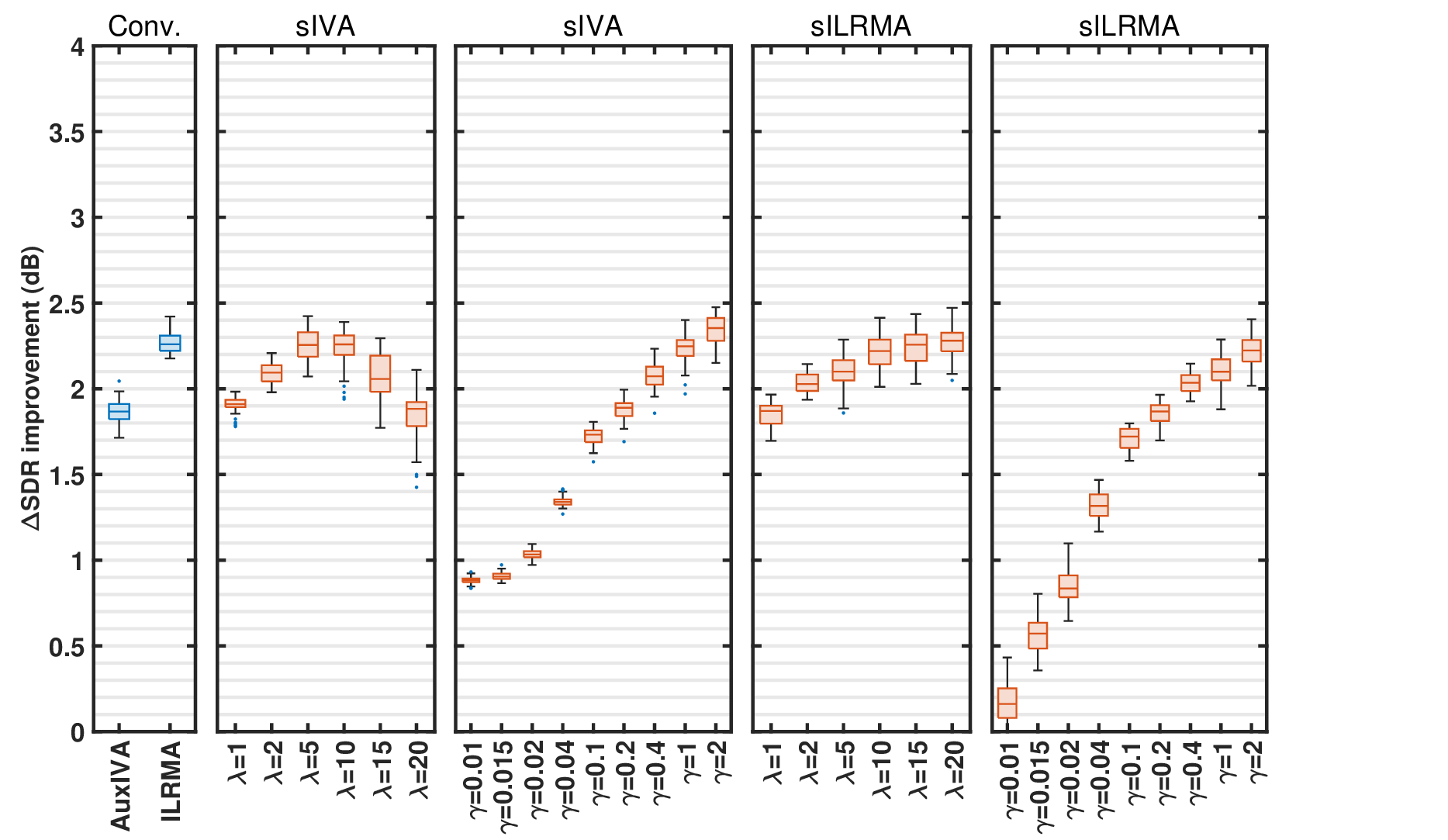}%
\label{Reuters_delta_SC}}\hspace{-1mm}
\hfil
\vspace{0.1cm}
\caption{SDR improvements for the 2-channel mixtures with the number of iterations being 100. The central lines indicate the median, the bottom and the top edges of the box indicate the 25th and 75th percentiles, respectively.}
\vspace{-0cm}
\label{fig21}
\end{figure}

\begin{figure}[!t]
\vspace{0cm}
\begin{minipage}[b]{1\linewidth}
  \centering
  \centerline{\includegraphics[width=3.4in]{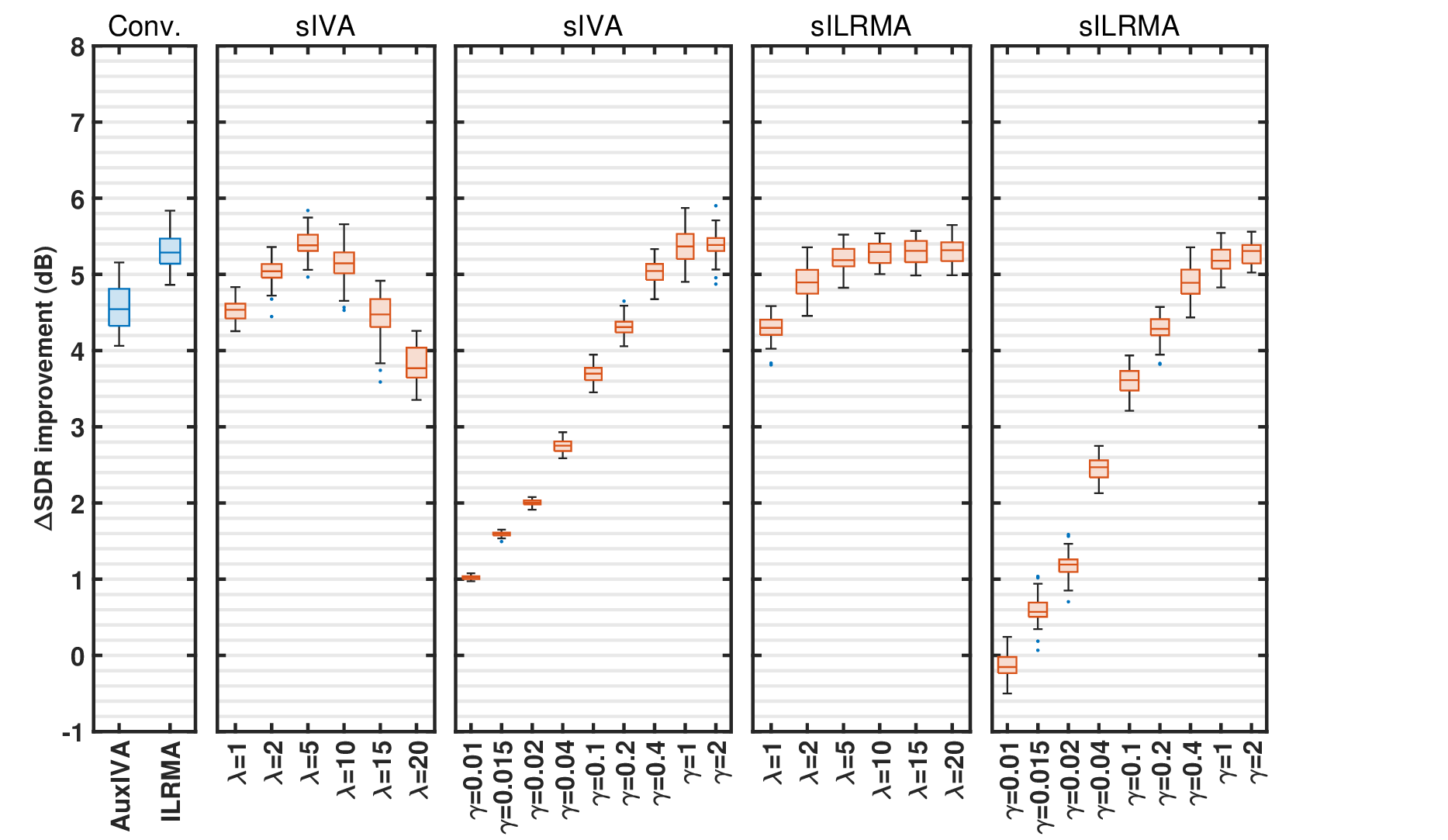}}
\end{minipage}
\vspace{-0.5cm}
\caption{SDR improvements for the 3-channel mixtures with the number of iterations being 100. The central lines indicate the median, the bottom and the top edges of the box indicate the 25th and 75th percentiles, respectively.}
\label{fig22}
\vspace{-0cm}
\end{figure}

\subsubsection{\fontperso{Scenario 2}}

To present the results for this scenario more concisely, the metrics have been averaged over different combinations and reported in Table~\ref{tab1} (Condition 1) and Table~\ref{tab2} (Condition 2). The proposed sILRMA achieves an SDR improvement of 4.5 dB compared to ILRMA in the anechoic environment. In the presence of reverberation, although the performance gap between sILRMA and ILRMA decreases as reverberation time increases, sILRMA consistently outperforms ILRMA. Additionally, sILRMA significantly surpasses ILRMA, GGDILRMA, tILRMA, and mILRMA across other evaluation metrics. These findings are confirmed by the average rank achieved by each method. Similar conclusions can be drawn from the results reported in Table~\ref{tab2} under Condition 2.

\begin{figure}[!t]
\vspace{-0cm} 
\setlength{\abovecaptionskip}{0cm} 
\setlength{\belowcaptionskip}{-0cm} 
\centering
\subfloat[\fontpersosmall{E2A}]{\includegraphics[width=3.4in]{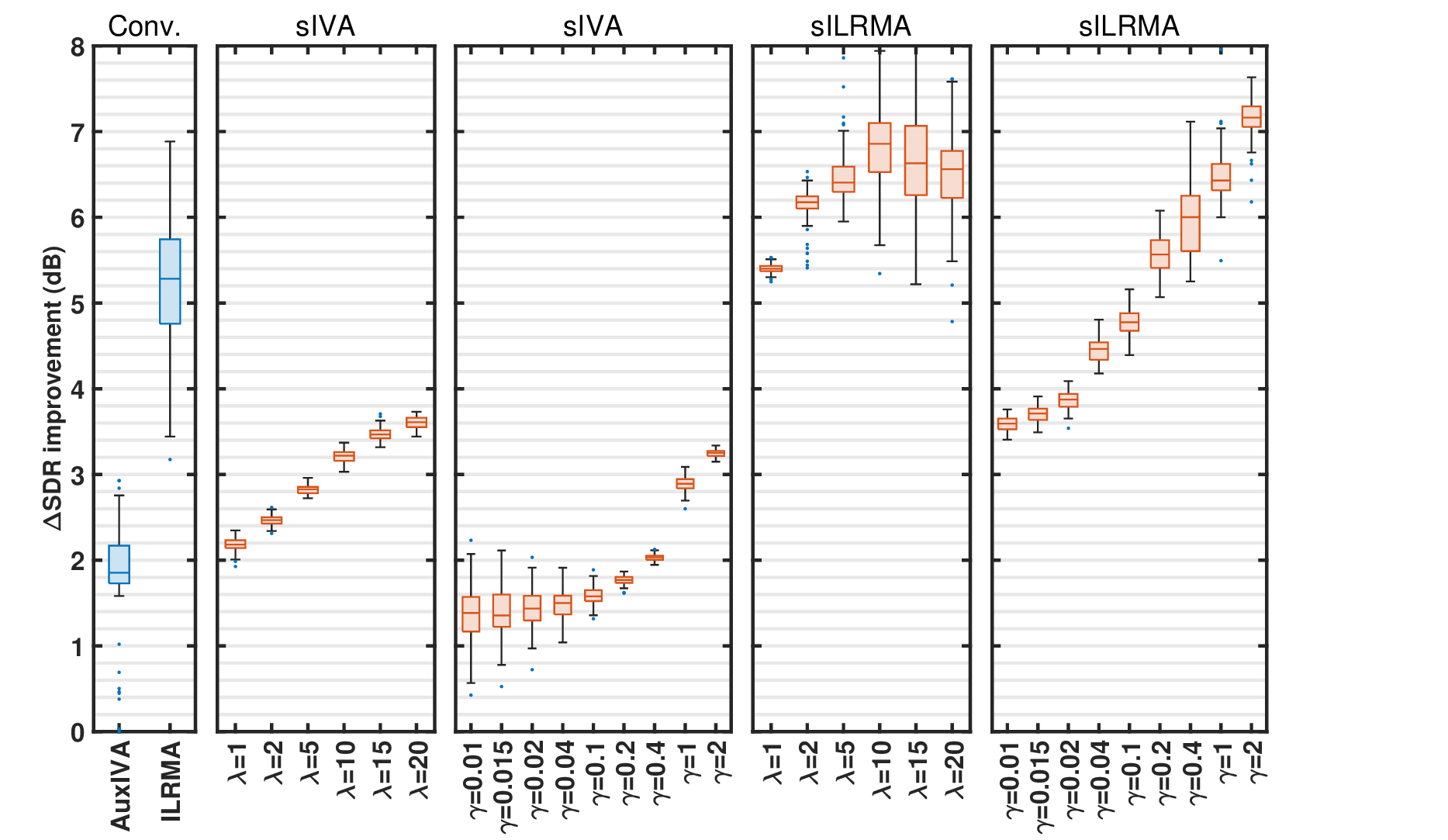}%
\label{Reuters_delta_Coh_E2A_2channels}}\hspace{-1mm}
\hfil
\subfloat[\fontpersosmall{JR2}]{\includegraphics[width=3.4in]{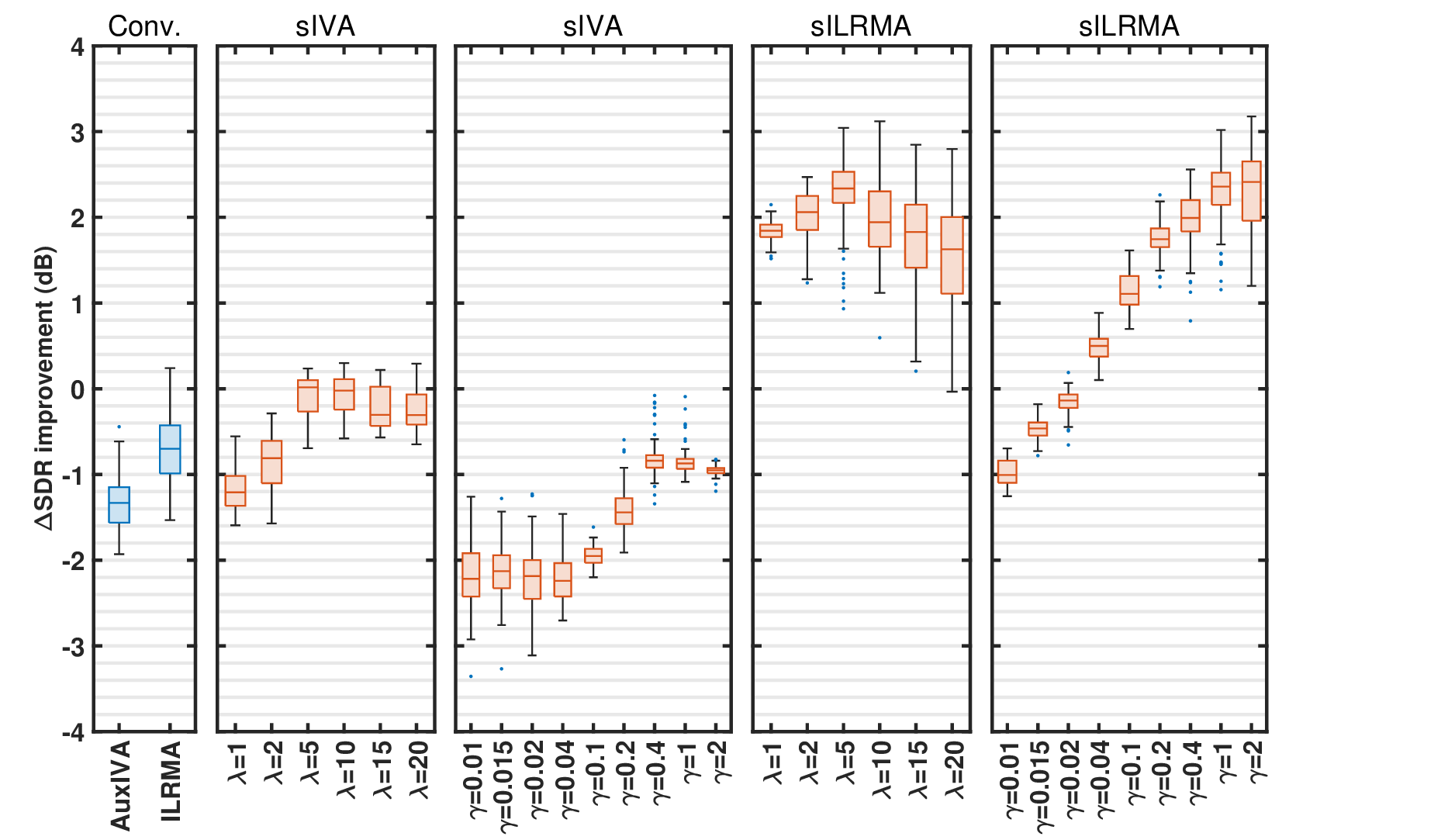}%
\label{Reuters_delta_SC_JR2_2channels}}\hspace{-1mm}
\hfil
\caption{SDR improvements for 2-channel mixtures with music signals and the number of iterations being 100. The central lines indicate the median, the bottom and the top edges of the box indicate the 25th and 75th percentiles, respectively.}
\vspace{-0cm}
\label{fig23}
\end{figure}

\begin{figure}[!t]
\vspace{-0cm} 
\setlength{\abovecaptionskip}{0cm} 
\setlength{\belowcaptionskip}{-0cm} 
\centering
\subfloat[E2A]{\includegraphics[width=3.4in]{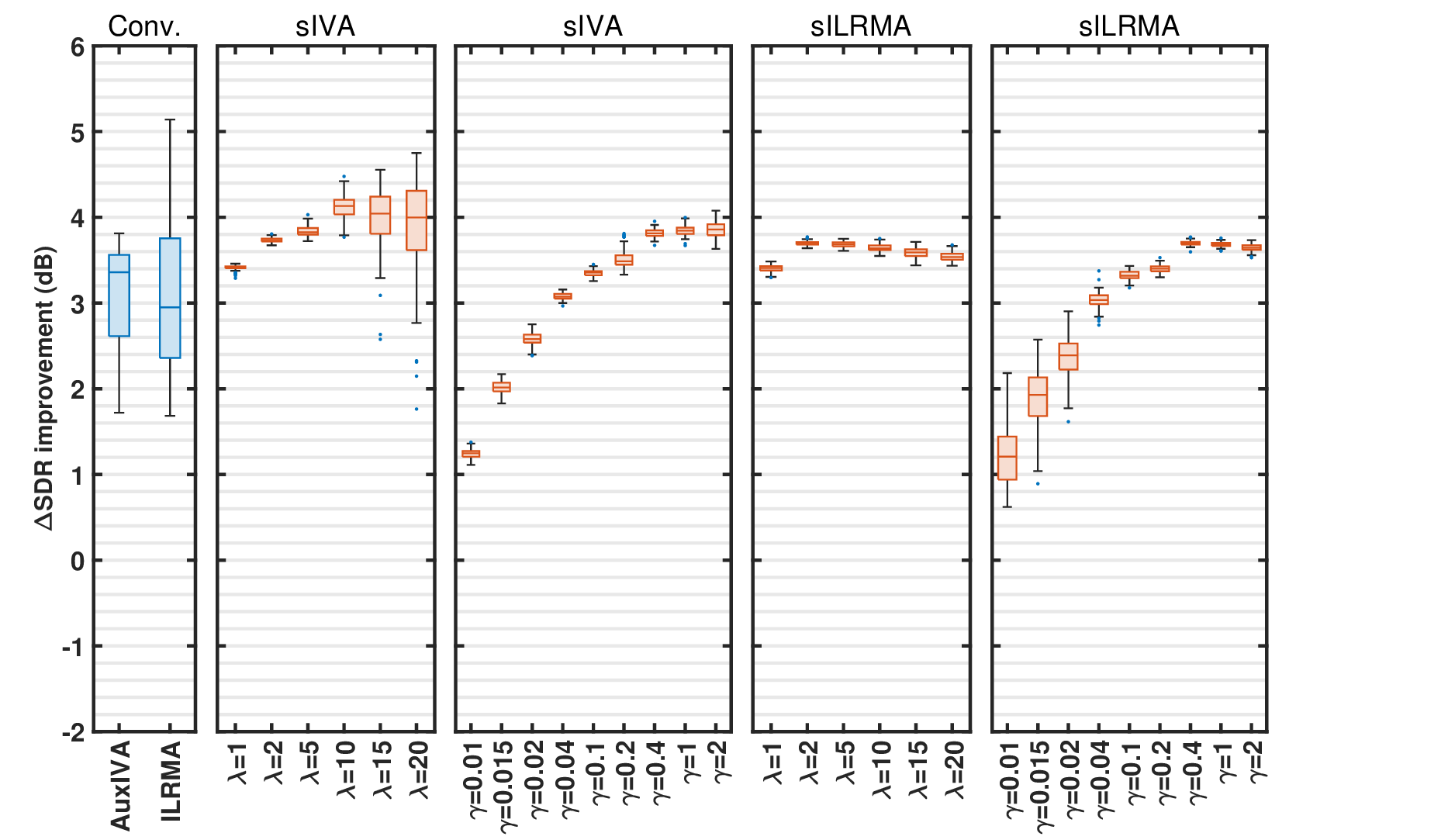}%
\label{Reuters_delta_Coh_E2A_3channels}}\hspace{-1mm}
\hfil
\subfloat[JR2]{\includegraphics[width=3.4in]{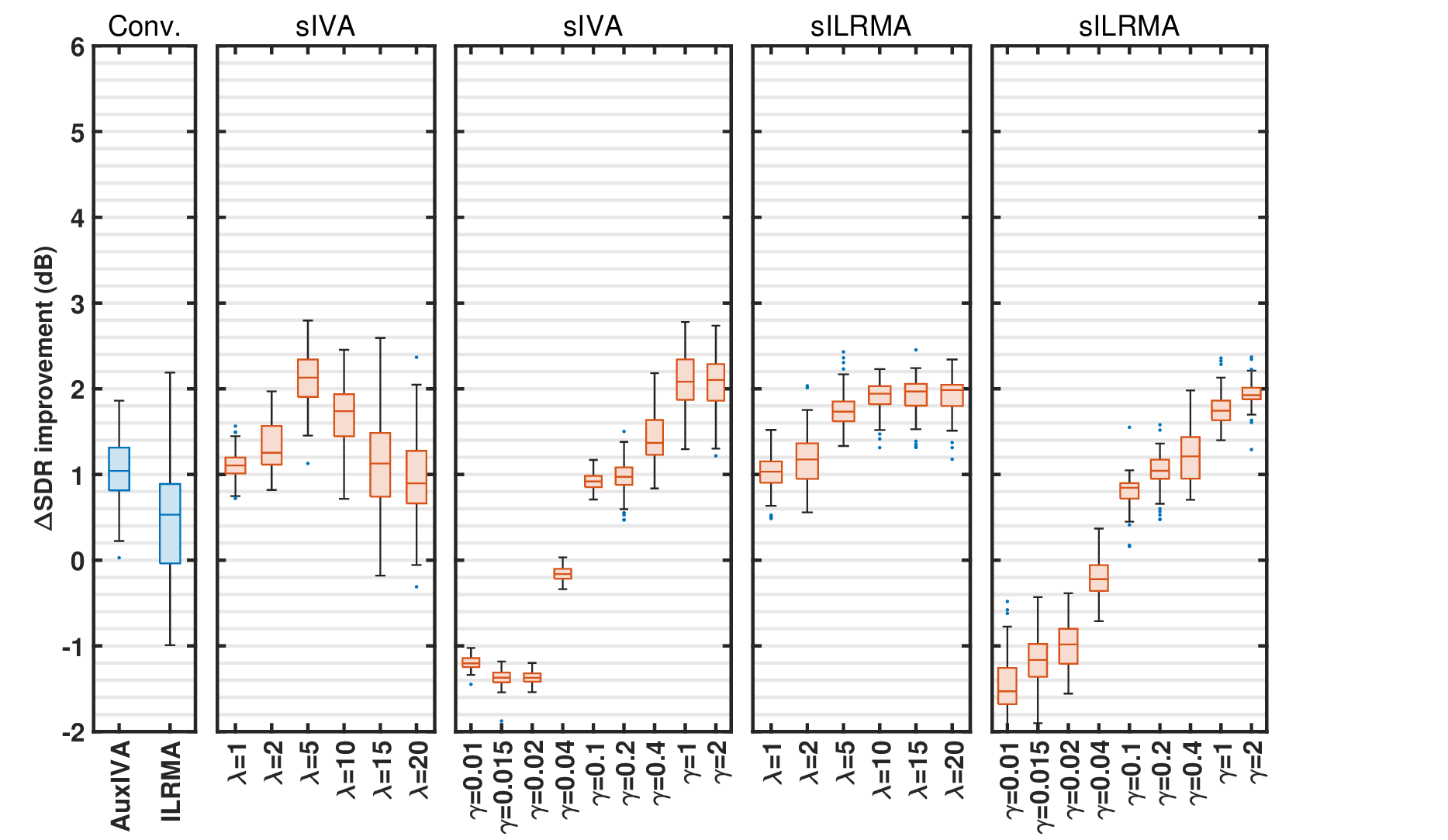}%
\label{Reuters_delta_SC_JR2_3channels}}\hspace{-1mm}
\hfil
\vspace{0.1cm}
\caption{SDR improvements for the 3-channel mixtures with music signals and the number of iterations being 100. The central lines indicate the median, the bottom and the top edges of the box indicate the 25th and 75th percentiles, respectively.}
\vspace{-0cm}
\label{fig24}
\end{figure}

\subsection{Sensitivity analysis}
To further validate the impact of algorithm hyperparameters on separation performance, additional experiments were conducted using a dataset provided by SiSEC for the underdetermined audio source separation task . The protocol from \cite{19} was followed, setting the frame length and overlap factor to 1024 points and $50\%$, respectively, in line with the methodology from \cite{16}.

To control variable influence when comparing algorithms, the same parameter initialization was used for AuxIVA, ILRMA, sIVA, and sILRMA. All algorithms underwent $100$ iterations. The number of components for ILRMA and sILRMA for each source was set to $K= 10$. The hyperparameter $\lambda$ for sIVA and sILRMA was chosen from $\{1,2,5,10,15,20\}$, while the hyperparameter $\gamma$ was fixed at 1. Conversely, for sIVA and sILRMA, the value of $\gamma$ was selected from $\{0.01, 0.015, 0.02, 0.04, 0.1, 0.2, 0.4, 1, 2\}$ with the value of $\lambda$ fixed at $5$.

The experimental results for the 2- and 3-channel cases are summarized in Figures~\ref{fig21} and \ref{fig22} for the SiSEC datasets referred to as \fontperso{dev1} and \fontperso{dev2}, respectively. On average, $\lambda=5$ and $\gamma = 1$ yield optimal performance for sIVA and sILRMA.

Furthermore, the experimental setup for music separation adheres to the setup outlined in \cite{19}. 2- and 3-channel mixtures were created by convolving the signals with the respective impulse responses \fontperso{E2A} and \fontperso{JR2} \cite{49}. From Figures~\ref{fig23} and \ref{fig24}, it is evident that the proposed algorithms achieve satisfactory results in the 2-channel scenario, although the improvement is less significant in the 3-channel case.

\section{Conclusion}
\label{Conclusion}
This paper addressed the challenge of leveraging inter-band dependencies and differences between distributions of source signals and their estimated counterparts to enhance traditional deterministic MBSS methods. Specifically, this work improved two popular deterministic BSS algorithms, IVA and ILRMA, with Sinkhorn divergence (a regularized form of the Wasserstein distance) to create improved variants of these algorithms. Integrating Sinkhorn divergence into MBSS not only facilitated modeling of inter-band signal dependencies but also aided in aligning estimated source signal distributions more closely with true source distributions. The effectiveness of these enhanced algorithms, i.e., sIVA and sILRMA, was rigorously evaluated through comprehensive simulations and experiments. These simulations and experiments included a detailed comparison involving six representative methods on both the simulated dataset WSJ0 and the real dataset {{SiSEC}}2011. The results demonstrated significant performance improvements of sIVA and sILRMA over the compared baseline methods across multiple evaluation metrics such as SDR, SIR, and SAR. This underscores the enhanced modeling capability of Sinkhorn divergence based approaches compared to traditional counterparts.

\bibliographystyle{IEEEbib}

\begin{thebibliography}{}

\end{thebibliography}


\begin{thebibliography}{10}

\bibitem{1}
L.~Parra and C.~Spence,
\newblock ``Convolutive blind separation of non-stationary sources,''
\newblock {\em IEEE Trans. Audio, Speech, Lang. Process.}, vol. 8, no. 3, pp.
  320--327, May 2000.

\bibitem{2}
J.~Benesty, S.~Makino, and J.~Chen,
\newblock {\em Speech Enhancement},
\newblock Springer, 2005.

\bibitem{3}
S.~Makino, T.W. Lee, and H.~Sawada,
\newblock {\em Blind speech separation},
\newblock Springer Dordrecht, 2007.

\bibitem{4}
D.~Kitamura, N.~Ono, H.~Sawada, H.~Kameoka, and H.~Saruwatari,
\newblock ``Determined blind source separation unifying independent vector
  analysis and nonnegative matrix factorization,''
\newblock {\em IEEE Trans. Audio, Speech, Lang. Process.}, vol. 24, no. 9, pp.
  1626--1641, Sept. 2016.

\bibitem{5}
P.~Bofill and M.~Zibulevsky,
\newblock ``Underdetermined blind source separation using sparse
  representations,''
\newblock {\em Signal Process.}, vol. 81, no. 11, pp. 2353--2362, Nov. 2001.

\bibitem{6}
P.~Comon,
\newblock ``Independent component analysis, a new concept?,''
\newblock {\em Signal Process.}, vol. 36, no. 3, pp. 287--314, Apr. 1994.

\bibitem{7}
P.~Smaragdis,
\newblock ``Blind separation of convolved mixtures in the frequency domain,''
\newblock {\em Neurocomputing}, vol. 22, no. 1-3, pp. 21--34, Nov. 1998.

\bibitem{8}
T.~Kim, T.~Eltoft, and T.W. Lee,
\newblock ``Independent vector analysis: An extension of ica to multivariate
  components,''
\newblock in {\em ICA}. Springer Berlin Heidelberg, 2006, pp. 165--172.

\bibitem{9}
T.~Kim, H.~T. Attias, S.Y. Lee, and T.W. Lee,
\newblock ``Blind source separation exploiting higher-order frequency
  dependencies,''
\newblock {\em IEEE Trans. Audio, Speech, Lang. Process.}, vol. 15, no. 1, pp.
  70--79, Jan. 2007.

\bibitem{10}
N.~Ono,
\newblock ``Stable and fast update rules for independent vector analysis based
  on auxiliary function technique,''
\newblock in {\em Proc. Int. Conf. Independent Compon. Anal. Blind Source
  Separation}. IEEE, Oct. 2011, pp. 189--192.

\bibitem{11}
D.~Kitamura, N.~Ono, H.~Sawada, H.~Kameoka, and H.~Saruwatari,
\newblock ``Determined blind source separation with independent low-rank matrix
  analysis,''
\newblock in {\em Audio Source Separation}, S.~Makino, Ed., pp. 125--155.
  Springer International Publishing, 2018.

\bibitem{14}
S.~Mogami, D.~Kitamura, Y.~Mitsui, N.~Takamune, H.~Saruwatari, and N.~Ono,
\newblock ``Independent low-rank matrix analysis based on complex student's
  t-distribution for blind audio source separation,''
\newblock in {\em 27th IEEE Int. Workshop Mach. Learn Signal Process. (MLSP)}.
  IEEE, Sept. 2017, pp. 1--6.

\bibitem{13}
K.~Kitamura, Y.~Bando, K.~Itoyama, and K.~Yoshii,
\newblock ``Student's t multichannel nonnegative matrix factorization for blind
  source separation,''
\newblock in {\em IWAENC}. IEEE, Sept. 2016, pp. 1--5.

\bibitem{15}
D.~Kitamura, S.~Mogami, Y.~Mitsui, N.~Takamune, H.~Saruwatari, N.~Ono,
  Y.~Takahashi, and K.~Kondo,
\newblock ``Generalized independent low-rank matrix analysis using heavy-tailed
  distributions for blind source separation,''
\newblock {\em EURASIP J. Adv. Signal Process.}, vol. 2018, no. 28, pp. 1--25,
  Apr. 2018.

\bibitem{33}
S.~Mogami, N.~Takamune, D.~Kitamura, H.~Saruwatari, Y.~Takahashi, K.~Kondo, and
  N.~Ono,
\newblock ``Independent low-rank matrix analysis based on time-variant
  sub-gaussian source model for determined blind source separation,''
\newblock {\em IEEE/ACM Trans. Audio, Speech, Language Process.}, vol. 28, pp.
  503--518, Dec. 2020.

\bibitem{22}
M.~Fontaine, K.~Sekiguchi, A.~A. Nugraha, Y.~Bando, and K.~Yoshii,
\newblock ``Generalized fast multichannel nonnegative matrix factorization
  based on gaussian scale mixtures for blind source separation,''
\newblock {\em IEEE/ACM Trans. Audio, Speech, Language Process.}, vol. 30, pp.
  1734--1748, May 2022.

\bibitem{18}
Y.~Mitsui, D.~Kitamura, S.~Takamichi, N.~Ono, and H.~Saruwatari,
\newblock ``Blind source separation based on independent low-rank matrix
  analysis with sparse regularization for time-series activity,''
\newblock in {\em Proc. IEEE Int. Conf. Acoust. Speech Signal Process.} IEEE,
  Mar. 2017, pp. 21--25.

\bibitem{16}
J.~Wang, S.~Guan, S.~Liu, and X.L. Zhang,
\newblock ``Minimum-volume multichannel nonnegative matrix factorization for
  blind audio source separation,''
\newblock {\em IEEE/ACM Trans. Audio, Speech, Language Process.}, vol. 29, pp.
  3089--3103, Oct. 2021.

\bibitem{17}
J.~Wang, S.~Guan, and X.L. Zhang,
\newblock ``Minimum-volume regularized ilrma for blind audio source
  separation,''
\newblock in {\em 2021 APSIPA ASC}, Dec. 2021, pp. 630--634.

\bibitem{19}
K.~Yatabe and D.~Kitamura,
\newblock ``Determined bss based on time-frequency masking and its application
  to harmonic vector analysis,''
\newblock {\em IEEE/ACM Trans. Audio, Speech, Language Process.}, vol. 29, pp.
  1609--1625, Apr. 2021.

\bibitem{21}
Y.~Mitsufuji, N.~Takamune, S.~Koyama, and H.~Saruwatari,
\newblock ``Multichannel blind source separation based on
  evanescent-region-aware non-negative tensor factorization in spherical
  harmonic domain,''
\newblock {\em IEEE/ACM Trans. Audio, Speech, Language Process.}, vol. 29, pp.
  607--617, Dec. 2021.

\bibitem{31}
R.~Badeau and M.~D. Plumbley,
\newblock ``Multichannel high-resolution nmf for modeling convolutive mixtures
  of non-stationary signals in the time-frequency domain,''
\newblock {\em IEEE/ACM Trans. Audio, Speech, Language Process.}, vol. 22, no.
  11, pp. 1670--1680, July 2014.

\bibitem{32}
T.~T.~H. Duong, N.~Q.~K. Duong, P.~C. Nguyen, and C.~Q. Nguyen,
\newblock ``Gaussian modeling-based multichannel audio source separation
  exploiting generic source spectral model,''
\newblock {\em IEEE/ACM Trans. Audio, Speech, Language Process.}, vol. 27, no.
  1, pp. 32--43, Sept. 2019.

\bibitem{36}
A.~Brendel, T.~Haubner, and W.~Kellermann,
\newblock ``A unified probabilistic view on spatially informed source
  separation and extraction based on independent vector analysis,''
\newblock {\em IEEE Trans. Signal Process.}, vol. 68, pp. 3545--3558, June
  2020.

\bibitem{37}
A.~Liutkus, R.~Badeau, and G.~Richard,
\newblock ``Gaussian processes for underdetermined source separation,''
\newblock {\em IEEE Trans. Signal Process.}, vol. 59, no. 7, pp. 3155--3167,
  July 2011.

\bibitem{38}
R.~Scheibler,
\newblock ``Independent vector analysis via log-quadratically penalized
  quadratic minimization,''
\newblock {\em IEEE Trans. Signal Process.}, vol. 69, pp. 2509--2524, Apr.
  2021.

\bibitem{sawada2023multi}
H.~Sawada, R.~Ikeshita, K.~Kinoshita, and T.~Nakatani,
\newblock ``Multi-frame full-rank spatial covariance analysis for
  underdetermined blind source separation and dereverberation,''
\newblock {\em IEEE/ACM Trans. Audio, Speech, Lang. Process.}, vol. 31, pp.
  3589--3602, 2023.

\bibitem{26}
J.~Nikunen and T.~Virtanen,
\newblock ``Direction of arrival based spatial covariance model for blind sound
  source separation,''
\newblock {\em IEEE/ACM Trans. Audio, Speech, Language Process.}, vol. 22, no.
  3, pp. 727--739, Jan. 2014.

\bibitem{23}
J.~J. Carabias-Orti, J.~Nikunen, T.~Virtanen, and P.~Vera-Candeas,
\newblock ``Multichannel blind sound source separation using spatial covariance
  model with level and time differences and nonnegative matrix factorization,''
\newblock {\em IEEE/ACM Trans. Audio, Speech, Language Process.}, vol. 26, no.
  9, pp. 1512--1527, Apr. 2018.

\bibitem{20}
K.~Sekiguchi, Y.~Bando, A.~A. Nugraha, K.~Yoshii, and T.~Kawahara,
\newblock ``Fast multichannel nonnegative matrix factorization with
  directivity-aware jointly-diagonalizable spatial covariance matrices for
  blind source separation,''
\newblock {\em IEEE/ACM Trans. Audio, Speech, Language Process.}, vol. 28, pp.
  2610--2625, Aug. 2020.

\bibitem{28}
N.~Ito, R.~Ikeshita, H.~Sawada, and T.~Nakatani,
\newblock ``A joint diagonalization based efficient approach to underdetermined
  blind audio source separation using the multichannel wiener filter,''
\newblock {\em IEEE/ACM Trans. Audio, Speech, Language Process.}, vol. 29, pp.
  1950--1965, May 2021.

\bibitem{34}
Y.~Mitsufuji, S.~Uhlich, N.~Takamune, D.~Kitamura, S.~Koyama, and
  H.~Saruwatari,
\newblock ``Multichannel non-negative matrix factorization using banded spatial
  covariance matrices in wavenumber domain,''
\newblock {\em IEEE/ACM Trans. Audio, Speech, Language Process.}, vol. 28, pp.
  49--60, Oct. 2020.

\bibitem{nakatani2022switching}
T.~Nakatani, R.~Ikeshita, K.~Kinoshita, H.~Sawada, N.~Kamo, and S.~Araki,
\newblock ``Switching independent vector analysis and its extension to blind
  and spatially guided convolutional beamforming algorithms,''
\newblock {\em IEEE/ACM Trans. Audio, Speech, Lang. Process.}, vol. 30, pp.
  1032--1047, 2022.

\bibitem{sekiguchi2022autoregressive}
K.~Sekiguchi, Y.~Bando, A.~A. Nugraha, M.~Fontaine, K.~Yoshii, and T.~Kawahara,
\newblock ``Autoregressive moving average jointly-diagonalizable spatial
  covariance analysis for joint source separation and dereverberation,''
\newblock {\em IEEE/ACM Trans. Audio, Speech, Lang. Process.}, vol. 30, pp.
  2368--2382, 2022.

\bibitem{5742773}
E.~Plourde and B.~Champagne,
\newblock ``Multidimensional stsa estimators for speech enhancement with
  correlated spectral components,''
\newblock {\em IEEE Trans. Signal Process.}, vol. 59, no. 7, pp. 3013--3024,
  2011.

\bibitem{54}
J.~Benesty, J.~Chen, and E.~Habets,
\newblock {\em Speech enhancement in the STFT domain},
\newblock Springer Berlin, Heidelberg, 2011.

\bibitem{chen2012single}
J.~Chen and J.~Benesty,
\newblock ``Single-channel noise reduction in the stft domain based on the
  bifrequency spectrum,''
\newblock in {\em Proc. IEEE Int. Conf. Acoust. Speech Signal Process.} IEEE,
  2012, pp. 97--100.

\bibitem{Liang2014}
Y.~Liang, S.~M. Naqvi, W.~Wang, and J.~A. Chambers,
\newblock {\em Frequency Domain Blind Source Separation Based on Independent
  Vector Analysis with a Multivariate Generalized Gaussian Source Prior}, pp.
  131--150,
\newblock Springer Berlin Heidelberg, Berlin, Heidelberg, 2014.

\bibitem{39}
A.~Rolet, M.~Cuturi, and G.~Peyré,
\newblock ``Fast dictionary learning with a smoothed wasserstein loss,''
\newblock in {\em AISTATS}. PMLR, May 2016, pp. 630--638.

\bibitem{41}
Kantorovich~L. V.,
\newblock ``On a problem of monge,''
\newblock {\em J. Math. Sci. (N.Y.)}, vol. 133, pp. 1383, 2006.

\bibitem{42}
S.~S. Vallender,
\newblock ``Calculation of the wasserstein distance between probability
  distributions on the line,''
\newblock {\em Theory Probab. its Appl.}, vol. 18, no. 4, pp. 784--786, 1974.

\bibitem{43}
C.~Frogner, C.~Zhang, H.~Mobahi, M.~Araya, and T.~A. Poggio,
\newblock ``Learning with a wasserstein loss,''
\newblock {\em Adv. Neural Inf. Process. Syst.}, vol. 28, pp. 2053--2061, Dec.
  2015.

\bibitem{44}
M.~Cuturi,
\newblock ``Sinkhorn distances: Lightspeed computation of optimal transport,''
\newblock {\em Adv. Neural Inf. Process. Syst.}, vol. 26, pp. 2292--2300, Dec.
  2013.

\bibitem{45}
E.~Cazelles, A.~Robert, and F.~Tobar,
\newblock ``The {W}asserstein-{F}ourier distance for stationary time series,''
\newblock {\em IEEE Trans. Signal Process.}, vol. 69, pp. 709--721, Dec. 2021.

\bibitem{fevotte2009nonnegative}
C.~F{\'e}votte, N.~Bertin, and JL. Durrieu,
\newblock ``Nonnegative matrix factorization with the itakura-saito divergence:
  With application to music analysis,''
\newblock {\em Neural Comput.}, vol. 21, no. 3, pp. 793--830, 2009.

\bibitem{lee2000algorithms}
D.~Lee and H.~S. Seung,
\newblock ``Algorithms for non-negative matrix factorization,''
\newblock {\em Adv. Neural Inf. Process.}, vol. 13, 2000.

\bibitem{50}
J.~Garofolo, D.~Graff, D.~Paul, and D.~Pallett,
\newblock ``Csr-i (wsj0) complete ldc93s6a,''
\newblock {\em Web Download. Philadelphia: Linguistic Data Consortium}, vol.
  83, 1993.

\bibitem{48}
S.~Araki, F.~Nesta, E.~Vincent, Z.~Koldovsk{\`y}, G.~Nolte, A.~Ziehe, and
  A.~Benichoux,
\newblock ``The 2011 signal separation evaluation campaign (sisec2011): - audio
  source separation -,''
\newblock in {\em LVA/ICA}. Springer Berlin Heidelberg, 2012, pp. 414--422.

\bibitem{51}
J.~B. Allen and D.~A. Berkley,
\newblock ``Image method for efficiently simulating small‐room acoustics,''
\newblock {\em J. Acoust. Soc. Am.}, vol. 65, no. 4, pp. 943--950, Jun 1979.

\bibitem{young1959sabine}
R.~W. Young,
\newblock ``Sabine reverberation equation and sound power calculations,''
\newblock {\em J. Acoust. Soc. Am.}, vol. 31, no. 7, pp. 912--921, 1959.

\bibitem{wang2024multichannel}
J.~Wang and S.~Guan,
\newblock ``Multichannel blind speech source separation with a disjoint
  constraint source model,''
\newblock {\em arXiv preprint arXiv:2401.01763}, 2024.

\bibitem{vincent2006performance}
E.~Vincent, R.~Gribonval, and C.~F{\'e}votte,
\newblock ``Performance measurement in blind audio source separation,''
\newblock {\em IEEE/ACM Trans. Audio, Speech, Language Process.}, vol. 14, no.
  4, pp. 1462--1469, 2006.

\bibitem{dubey2023icassp}
H.~Dubey, A.~Aazami, V.~Gopal, B.~Naderi, S.~Braun, R.~Cutler, H.~Gamper, M.~Golestaneh, and R.~Aichner,
\newblock ``Icassp 2023 deep noise suppression challenge,''
\newblock in {\em ICASSP}, 2023.

\bibitem{49}
S.~Nakamura, K.~Hiyane, F.~Asano, T.~Nishiura, and T.~Yamada,
\newblock ``Acoustical sound database in real environments for sound scene
  understanding and hands-free speech recognition.,''
\newblock {\em Proc. Int. Conf. Lang. Resour. Eval.}, pp. 965--968, June 2000.

\bibitem{peyre2019computational}
G.~Peyr{\'e}, and M.~Cuturi,
\newblock ``Computational optimal transport: With applications to data science,''
\newblock {Foundations and Trends{\textregistered} in Machine Learning}, 2019.

\bibitem{scheibler2019independent}
R.~Scheibler, and N.~Ono,
\newblock ``Independent vector analysis with more microphones than sources,''
\newblock {IEEE Worksh. Appl. Signal Process. Audio Acoust.}, pp. 185--189, 2019.

\bibitem{scheibler2019independent}
A.~Rolet, V.~Seguy, M.~Blondel, and H.~Sawada,
\newblock ``Blind source separation with optimal transport non-negative matrix factorization,''
\newblock {EURASIP J. Adv. Signal Process.}, pp. 1--16, 2018.


\end{thebibliography}

\end{document}